\documentclass[12pt]{article}

\usepackage[dvipdfmx]{graphicx}
\usepackage{tikz}
\usetikzlibrary{decorations.pathmorphing}

\usepackage{fancybox}

\usepackage{cite}
\usepackage{float}
\usepackage{amsfonts}
\usepackage{amsmath}
\usepackage{amsbsy}
\usepackage{graphicx}
\usepackage{amssymb}
\usepackage{amsthm}
\usepackage{bm}
\usepackage{epsfig}
\usepackage{latexsym}
\usepackage{pdflscape}
\usepackage{color}
\usepackage{here}
\usepackage{graphicx}
\numberwithin{equation}{section}

\allowdisplaybreaks

\newcounter{aff}

\begin{document}
\begin{titlepage}

\begin{flushright}
{\footnotesize OCU-PHYS 467, KIAS-P17046}
\end{flushright}
\begin{center}
{\LARGE\bf 
Superconformal Chern-Simons Theories\\[6pt]
from del Pezzo Geometries
}\\
\bigskip\bigskip
{\large 
Sanefumi Moriyama\,\footnote{\tt moriyama@sci.osaka-cu.ac.jp},
\quad
Tomoki Nosaka\,\footnote{\tt nosaka@yukawa.kyoto-u.ac.jp},
\quad
Katsuya Yano\,\footnote{\tt yanok@sci.osaka-cu.ac.jp}
}\\
\bigskip\bigskip
${}^{*\ddagger}$\,{\it Department of Physics, Graduate School of Science,
Osaka City University}\\
${}^*$\,{\it Osaka City University 
Advanced Mathematical Institute (OCAMI)}\\
{\it Sumiyoshi-ku, Osaka 558-8585, Japan}\\[6pt]
${}^\dagger$\,{\it Korea Institute for Advanced Study}\\
{\it Dongdaemun-gu, Seoul 02455, Korea}
\end{center}

\begin{abstract}
We present an explicit expression for the grand potential of the U$(N)^3$ superconformal Chern-Simons theory with the Chern-Simons levels being $(k,0,-k)$.
From the viewpoint of the Newton polygon, it is expected that the grand potential is given by the free energy of the topological string theory on the local $D_5$ del Pezzo geometry, though the explicit identification was a puzzle for years.
We show how the expectation is realized explicitly.
As a bonus, we can also study the ${\mathbb Z}_2$ orbifold of this theory and find the grand potential is now given in terms of the local $E_7$ del Pezzo geometry.
\end{abstract}

\end{titlepage}
\setcounter{footnote}{0}

\tableofcontents

\section{Introduction}

M-theory, though it was proposed to unify all of the five perturbative string theories, has been a mysterious theory for a long time.
Recently this theory was demystified largely partially due to the discovery of the worldvolume theory of the fundamental M2-branes.
Namely, it was proposed \cite{ABJM,HLLLP2,ABJ} that the worldvolume theory of $\min(N_1,N_2)$ M2-branes and $|N_2-N_1|$ fractional M2-branes on the target space geometry ${\mathbb C}^4/{\mathbb Z}_k$ is described by the ${\mathcal N}=6$ superconformal Chern-Simons theory with the gauge group U$(N_1)_k\times$U$(N_2)_{-k}$ and two pairs of bifundamental matters where the subscripts $(k,-k)$ denote the Chern-Simons levels. 

Due to the localization techniques \cite{Pestun,KWY}, the infinite-dimensional path integral in defining the partition function of the ABJM theory on $S^3$ is reduced to a finite-dimensional matrix integration.
It is convenient to consider the reduced grand potential\footnote{See \eqref{definitionofJ} later for the definition of the reduced grand potential.} $J(\mu)$ \cite{HMO2} for the partition function by regarding the rank $N=\min(N_1,N_2)$ as the number of particles and introducing the dual chemical potential $\mu$ \cite{MP}.
Then, it was known \cite{HMMO}\footnote{See also \cite{DMP1,HKPT,DMP2,FHM,MP,HMO1,PY,HMO2,CM,HMO3} for earlier works leading to this result.} that, if we further redefine the effective chemical potential $\mu_\text{eff}$ appropriately \cite{HMO3}, aside from the perturbative part of the reduced grand potential given by a cubic polynomial of the effective chemical potential \cite{DMP1,FHM,MP}, the non-perturbative part is separated into that of pure worldsheet instantons \cite{DMP1,CSW} and that of pure membrane instantons \cite{DMP2}, $J^\text{np}(\mu_\text{eff})=J^\text{WS}(\mu_\text{eff})+J^\text{MB}(\mu_\text{eff})$.
The worldsheet instanton $J^\text{WS}(\mu_\text{eff})$ takes the form of the free energy of the topological string theory, while the membrane instanton $J^\text{MB}(\mu_\text{eff})$ takes the form of the derivative of the free energy of the refined topological string theory in the Nekrasov-Shatashvili limit ($s_{\text{L}/\text{R}}=2j_{\text{L}/\text{R}}+1$)
\begin{align}
J^\text{WS}(\mu_\text{eff})&=\sum_{j_\text{L},j_\text{R}}\sum_{\bm d}N^{\bm d}_{j_\text{L},j_\text{R}}
\sum_{n=1}^\infty\frac{(-1)^{(s_\text{L}+s_\text{R}-1)n}s_\text{R}\sin 2\pi g_\text{s}ns_\text{L}}
{n(2\sin\pi g_\text{s}n)^2\sin 2\pi g_\text{s}n}e^{-n{\bm d}\cdot{\bm T}},
\nonumber\\
J^\text{MB}(\mu_\text{eff})&=\sum_{j_\text{L},j_\text{R}}\sum_{\bm d}N^{\bm d}_{j_\text{L},j_\text{R}}
\sum_{n=1}^\infty\frac{\partial}{\partial g_\text{s}}
\biggl[g_\text{s}\frac{-\sin\frac{\pi n}{g_\text{s}}s_\text{L}\sin\frac{\pi n}{g_\text{s}}s_\text{R}}
{4\pi n^2(\sin\frac{\pi n}{g_\text{s}})^3}e^{-n\frac{{\bm d}\cdot{\bm T}}{g_\text{s}}}\biggr].
\label{np}
\end{align}
Here the two K\"ahler parameters and the string coupling constant are identified as
\begin{align}
T^\pm=\frac{4\mu_\text{eff}}{k}\pm\pi i\biggl(1-\frac{2M}{k}\biggr),\quad
g_\text{s}=\frac{2}{k},
\end{align}
with $M=N_2-N_1$ and $N^{\bm d}_{j_\text{L},j_\text{R}}$ is the BPS indices of the local ${\mathbb P}^1\times{\mathbb P}^1$ geometry (see \cite{PTEP,Marino} for reviews).
The appearance of the topological string theory and the local ${\mathbb P}^1\times{\mathbb P}^1$ geometry may look surprising at first sight.
This is partially motivated by the Fermi gas formalism \cite{MP}, which rewrites the partition function of the ABJM theory into that of a non-interacting Fermi gas system.
The spectral operator of this system is given by $e^{\widehat H}=(2\cosh\frac{\widehat q}{2})(2\cosh\frac{\widehat p}{2})$ where ${\widehat q}$ and ${\widehat p}$ are the canonical position/momentum operators.
Then, it was observed \cite{MP} that the Newton polygon of the classical spectral curve $\sum_{m,n}e^{mq+np}=e^E$ with $m,n=\pm\frac{1}{2}$ is nothing but that of the ${\mathbb P}^1\times{\mathbb P}^1$ geometry under the change of variables.

After establishing the results for the M2-branes on the background with large supersymmetry, it is interesting to explore more general backgrounds.
Namely, we can naturally ask what happens when we consider other superconformal Chern-Simons theories, which are natural generalizations of the ABJM theory.
Especially, we are interested in whether the non-perturbative part of the reduced grand potential of those superconformal Chern-Simons theories falls into the same expression \eqref{np}, or if not, what the generalization of \eqref{np} is.
Interestingly, in \cite{GHM1} it was conjectured that the reduced grand potential of a large class of the spectral determinants falls into the same expression as \eqref{np}, where the geometry is read off from the classical spectral curve as in the case of the ABJM theory.

The investigation of the grand potential of general superconformal Chern-Simons theories starts from a special class enjoying the supersymmetry ${\mathcal N}=4$.
It was found \cite{GW,HLLLP1,IK1,TY,IK2} that for the circular quiver of unitary gauge groups the superconformal Chern-Simons theory enjoys the supersymmetry enhancement of ${\mathcal N}=4$ if the Chern-Simons levels satisfies $k_a=(k/2)(s_a-s_{a-1})$ with $s_a=\pm 1$.

One of the simplest models \cite{MN3} among the ${\mathcal N}=4$ superconformal Chern-Simons theories is the theory with the gauge group U$(N)_k\times$U$(N)_{0}\times$U$(N)_{-k}\times$U$(N)_{0}$, which is dubbed $(2,2)$ model from the number of $\pm 1$ appearing continuously in $\{s_a\}=\{+1,+1,-1,-1\}$.
In fact, it was observed \cite{MN3} that the non-perturbative part of the grand potential has the structure of \eqref{np} with $g_s=1/k$ and a particular choice of K\"ahler parameters $T$.
Moreover, the diagonal Gopakumar-Vafa invariants, special combinations of the BPS indices, of the $(2,2)$ model match with those of the local $D_5$ del Pezzo geometry.
This is indeed natural from the viewpoint of the Newton polygon since the spectral curve of the $(2,2)$ model is $\sum_{m,n}e^{mq+np}=e^E$ with $m,n=0,\pm 1$.

Due to the complexity with large degrees of freedom, it was difficult to study this generalization carefully.
Very recently, from the improvements in the Fermi gas formalism,\footnote{See \cite{HHMO,MM,ADF,MN4,N,MS2,MN5,MaMo,KiMo} for related improvements in the Fermi gas formalism.} we were able to revisit the $(2,2)$ model by considering the rank deformations \cite{MNN} and found that the reduced grand potential of the rank deformed $(2,2)$ model still falls into the same non-perturbative expression \eqref{np} with the total BPS indices listed in \cite{HKP} split in a very non-trivial way.
We also studied the rank deformations of the ${\mathbb Z}_2$ orbifold\footnote{
The physical interpretation of the repetition of the spectral operator is the orbifold in the target space of the M2-branes \cite{HLLLP1,IK1}.
This should not be confused with the orbifold in the background geometry of the topological string theory.
}
of the ABJM theory, or the $(1,1,1,1)$ model with $\{s_a\}=\{+1,-1,+1,-1\}$, which are connected to the $(2,2)$ model at the edge of the rank deformations through the Hanany-Witten duality \cite{HW}.
We found that the free energy of the topological string theory \eqref{np} unifies the moduli space of the rank deformations of these two dual models with the six K\"{a}hler parameters of the local $D_5$ del Pezzo geometry.
From this unified viewpoint, the worldsheet instanton exponent $e^{-\frac{2\mu_\text{eff}}{k}}$ of the $(1,1,1,1)$ model is realized by a non-trivial cancellation in the worldsheet instantons whose exponent is generically $e^{-\frac{\mu_\text{eff}}{k}}$.

Another interesting model is the $(2,1)$ model with the gauge group U$(N)_k\times$U$(N)_{0}\times$U$(N)_{-k}$ whose levels are specified by $\{s_a\}=\{+1,+1,-1\}$.
Although the study of this model dates back to \cite{MN1}, it was, however, difficult to find the general structure for a long time.
In this paper, we shall present a complete description of the $(2,1)$ model (without rank deformations).
We have found that the description of the $(2,1)$ model falls into the same expression as \eqref{np} if we choose the K\"ahler parameters and the BPS indices appropriately, though it looks quite different at first sight.

The study of this model is interesting also from the viewpoint of the Newton polygon.
Though the spectral curve of the $(2,1)$ model is $\sum_{m,n}e^{mq+np}=e^E$ with $m=0,\pm 1$, $n=\pm\frac{1}{2}$, we cannot consistently truncate to these points in the Newton polygon.
In fact, after rescaling $p/2\to p$, the Newton polygon is indistinguishable as a convex hull from that for the $(2,2)$ model.
So our main task in this paper is to identify how the $D_5$ del Pezzo geometry appears in the $(2,1)$ model.
After observing that the instanton expression of the $(2,1)$ model keeps many BPS indices of the local $D_5$ del Pezzo geometry as mementos, we construct a framework so that these mementos can be utilized to describe the model correctly.

Considering the rather long analysis of the $(2,1)$ model starting from \cite{MN1}, our resulting statement is surprisingly short.
The reduced grand potential of the $(2,1)$ model is given by the same expression of topological strings \eqref{np} with the four K\"ahler parameters
\begin{align}
T_\uparrow^\pm=\frac{2\mu_\text{eff}}{k}\pm\pi i\biggl(1+\frac{1}{k}\biggr),\quad
T_\downarrow^\pm=\frac{2\mu_\text{eff}}{k}\pm\pi i\biggl(-1+\frac{1}{k}\biggr).
\end{align}
The BPS indices are obtained by identifying those of the local $D_5$ del Pezzo  geometry as the representations of the original algebra so$(10)$ and decomposing the representations to the subalgebra so$(6)\times$u$(1)\times$u$(1)$ where the two u$(1)$ charges are identified respectively as the two degree differences of $\pm$ and $\uparrow\downarrow$.

As a bonus of our study, we can also study the ${\mathbb Z}_2$ orbifold of the $(2,1)$ model, that is, the $(2,1,2,1)$ model with $\{s_a\}=\{+1,+1,-1,+1,+1,-1\}$.
We have identified the reduced grand potential of the $(2,1,2,1)$ model with the topological string description \eqref{np} with the BPS indices being those of the local $E_7$ del Pezzo geometry.
This is motivated by a suggestive expression of the Newton polygon of the $E_7$ del Pezzo geometry in \cite{KY}.

The organization of this paper is as follows.
In section \ref{21} we first review the known results of the $(2,1)$ model.
After acquiring some clues from the observations on the relation to the rank-deformed $(2,2)$ model with the gauge group U$(N)_k\times$U$(N+M)_0\times$U$(N+2M)_{-k}\times$U$(N+M)_0$ and on the group-theoretical viewpoint for the $(2,2)$ model in section \ref{obs}, in section \ref{top} we present carefully how the reduced grand potential is described with the free energy of topological strings.
In section \ref{22} we shortly revisit the two-parameter rank deformation of the $(2,2)$ model $\text{U}(N+M_\text{II})_k\times\text{U}(N+M_\text{I})_0\times\text{U}(N+2M_\text{I}+M_\text{II})_{-k}\times\text{U}(N+M_\text{I})_0$ studied in \cite{MNN} by expressing the reduced grand potential in a more economical language of characters.
In section \ref{2121} we turn to the $(2,1,2,1)$ model and describe the reduced grand potential of this model using the language of characters.
Finally we conclude with some discussions.
 
In appendix \ref{D5data} we summarize the instanton coefficients and the group-theoretical data which are necessary in order to check the relation between the representation theory for $\text{so}(10)$ and the instanton coefficients of the $(2,1)$ model and the rank deformed $(2,2)$ model.
Appendix \ref{E7data} is the collection of the instanton coefficients of the $(2,1,2,1)$ model and the group-theoretical data for $E_7$ and $\text{so}(12)$ relevant to our proposal.

\section{$(2,1)$ model}\label{21}

In this section we review the result for the $(2,1)$ model \cite{MN1,MN3} shortly.
The infinite-dimensional path integral in defining the partition function of the $(2,1)$ model is reduced to a finite-dimensional matrix integration \cite{KWY}
\begin{align}
Z(N)=\int\frac{D^N\mu}{N!}\frac{D^N\lambda}{N!}\frac{D^N\nu}{N!}
\frac{\prod_{m<m'}^N(2\sinh\frac{\mu_m-\mu_{m'}}{2})^2
\prod_{l<l'}^N(2\sinh\frac{\lambda_l-\lambda_{l'}}{2})^2
\prod_{n<n'}^N(2\sinh\frac{\nu_n-\nu_{n'}}{2})^2}
{\prod_{m,l}^N2\cosh\frac{\mu_m-\lambda_l}{2}
\prod_{l,n}^N2\cosh\frac{\lambda_l-\nu_n}{2}
\prod_{n,m}^N2\cosh\frac{\nu_n-\mu_m}{2}},
\end{align}
with the integrations
\begin{align}
D\mu=\frac{d\mu}{2\pi}e^{\frac{ik}{4\pi}\mu^2},\quad
D\lambda=\frac{d\lambda}{2\pi},\quad
D\nu=\frac{d\nu}{2\pi}e^{-\frac{ik}{4\pi}\nu^2}.
\end{align}
It was found that the {\it reduced} grand potential of the $(2,1)$ model defined as\footnote{
See \cite{HMO2} for an explanation on the reason to study the reduced grand potential instead of the original grand potential, which is defined simply as $e^{J^{\text{original}}(\mu)}$ for the same right-hand side of \eqref{definitionofJ}.
}
\begin{align}
\sum_{n=-\infty}^\infty e^{J(\mu+2\pi in)}
=\sum_{N=0}^\infty e^{N\mu}Z(N),
\label{definitionofJ}
\end{align}
by introducing the chemical potential $\mu$ dual to the rank $N$, is given separately as the summation of the worldsheet instanton part and the membrane instanton part
\begin{align}
J(\mu)=J^\text{pert}(\mu_\text{eff})+J^\text{np}(\mu_\text{eff}),\quad
J^\text{np}(\mu_\text{eff})=J^\text{WS}(\mu_\text{eff})+J^\text{MB}(\mu_\text{eff}),
\label{J21}
\end{align}
aside from the perturbative part,
\begin{align}
J^\text{pert}(\mu_\text{eff})=\frac{C}{3}\mu_\text{eff}^3+B\mu_\text{eff}+A,\quad
C=\frac{1}{\pi^2k},\quad
B=-\frac{1}{12k}+\frac{k}{12},
\label{21pert}
\end{align}
with $A$ given in \cite{MN1}, if we reexpress with the effective chemical potential $\mu_\text{eff}$ suitably.
For integral $k$, $\mu_\text{eff}$ is given by
\begin{align}
\mu_\text{eff}=\begin{cases}
\displaystyle
\mu-2e^{-2\mu}{}_4F_3\Bigl(1,1,\frac{3}{2},\frac{3}{2};
2,2,2;16e^{-2\mu}\Bigr),&\text{ for odd }k,\\[6pt]
\displaystyle
\mu-6e^{-2\mu}{}_4F_3\Bigl(1,1,\frac{7}{4},\frac{5}{4};
2,2,2;64e^{-2\mu}\Bigr),&\text{ for even }k,
\end{cases}
\label{mueff21}
\end{align}
where the first few non-perturbative terms are extrapolated into real functions of $k$ using the WKB expansion \cite{MN3}.

The worldsheet instantons are given by
\begin{align}
J^\text{WS}(\mu_\text{eff})
=\sum_{m=1}^\infty d_me^{-m\frac{2\mu_\text{eff}}{k}}.
\label{21ws}
\end{align}
The coefficients $d_m$ are determined as real functions of $k$ by the interpolation from the coefficients at integral $k$, which are found to satisfy the multi-covering structure
\begin{align}
d_m=\sum_{n|m}\frac{1}{n}\delta_{\frac{m}{n}}\biggl(\frac{k}{n}\biggr),
\label{wsmc}
\end{align}
where the multi-covering component $\delta_d(k)$ takes the following form
\begin{align}
\delta_d(k)=\frac{\sum_n\delta_{d,n}\cos\frac{\pi n}{k}}{(2\sin\frac{2\pi}{k})^2},
\end{align}
with a finite number of non-vanishing integral coefficients $\delta_{d,n}$ at each degree.
The first several components $\delta_d(k)$ are summarized in appendix \ref{21instanton}.

The membrane instantons are given by the general form
\begin{align}
J^\text{MB}(\mu_\text{eff})
=\widetilde J_b(\mu_\text{eff})\mu_\text{eff}+\widetilde J_c(\mu_\text{eff}),\quad
\widetilde J_b(\mu_\text{eff})
=\sum_{\ell=1}^\infty\widetilde b_{2\ell}e^{-2\ell\mu_\text{eff}},\quad
\widetilde J_c(\mu_\text{eff})
=\sum_{\ell=1}^\infty\widetilde c_\ell e^{-\ell\mu_\text{eff}},
\label{21mbexpand}
\end{align}
where the instanton coefficients of odd instantons $\widetilde c_{2\ell-1}$ are constants in $\mu_\text{eff}$, while those of even instantons are the standard linear polynomials in $\mu_\text{eff}$ with $\widetilde b_{2\ell}\mu_\text{eff}+\widetilde c_{2\ell}$ satisfying the derivative relation
\begin{align}
\widetilde c_{2\ell}=-k^2\frac{d}{dk}\frac{\widetilde b_{2\ell}}{2\ell k}.
\label{derivative}
\end{align}

The first several coefficients were investigated from the WKB expansion up to ${\mathcal O}(k^9)$ in \cite{MN1,MN3}.
The coefficients of the odd instantons can be expressed in the following simple multi-covering structure 
\begin{align}
{\widetilde c}_{2\ell-1}=\sum_{n|2\ell-1}\frac{(-1)^{\frac{n-1}{2}}}{n}
\gamma_{\frac{2\ell-1}{n}}(nk),
\label{coddmc}
\end{align}
or explicitly
\begin{align}
\widetilde c_1=\gamma_1(k),\quad
\widetilde c_3=-\frac{1}{3}\gamma_1(3k)+\gamma_3(k),\quad
\widetilde c_5=\frac{1}{5}\gamma_1(5k)+\gamma_5(k),\quad
\cdots,
\label{codd}
\end{align}
where $\gamma_d(k)$ takes the following form
\begin{align}
\gamma_d(k)=-\frac{\sum_n\gamma_{d,n}\sin\pi nk}{\sin^2\frac{\pi k}{2}},
\end{align}
with a finite number of positive integral coefficients $\gamma_{d,n}$ at each degree.
Once we accept this multi-covering structure and utilize the WKB expansion of surprisingly high order ${\cal O}(k^{29})$ \cite{HHO} obtained with the derivative formalism of \cite{MN2}, we can further determine ${\gamma}_d(k)$ of higher degree $d$.
The explicit expressions of the functions $\gamma_d(k)$ are listed in appendix \ref{21instanton}.

The multi-covering structure for the even instantons was not clearly understood.
Nevertheless, we achieved to determine the first few coefficients without recognizing the multi-covering structure
\begin{align}
{\widetilde b}_2&=\frac{8+11\cos\pi k+8\cos 2\pi k+\cos 3\pi k}{\pi\sin 2\pi k},\nonumber \\
{\widetilde b}_4&=\frac{136+256\cos \pi k+255\cos 2\pi k+192\cos 3\pi k+136\cos 4\pi k+64\cos 5\pi k+21\cos 6\pi k}{2\pi\sin 4\pi k},
\label{btildewithoutMC}
\end{align}
from the ansatz
\begin{align}
{\widetilde b}_{2\ell}=\frac{\sum_n{\widetilde b}_{2\ell,n}\cos \pi nk}{\ell\pi\sin 2\pi \ell k},
\end{align}
with a finite number of non-vanishing integers ${\widetilde b}_{2\ell,n}$.
With the abundant WKB data \cite{HHO} we could further determine higher instanton coefficients ${\widetilde b}_{2\ell}$.
Before going on to the higher instantons, however, let us provide several new observations which are essential to reveal the whole structure of the instanton coefficients.

\section{Observations}\label{obs}

In this section we shall make several observations for the non-perturbative part of the $(2,1)$ model and the $(2,2)$ model, which are helpful later in solving the models.

\subsection{Worldsheet instanton relation}\label{wsrel}

In \cite{MN3} we observed that when setting all the cosine functions in the numerators of the worldsheet instantons of the $(2,1)$ model in \eqref{21ws1to7} to be $1$ (with the replacement of $k$ by $2k$) we correctly reproduce the worldsheet instantons of the $(2,2)$ model for $1\le d\le 5$.
This relation is not valid any more for higher instantons, though the expressions look close.
We find that this observation should be replaced by the following more accurate observation.

In \cite{MNN} we studied the $(2,2)$ model with rank deformations.
Among others, it was found that the worldsheet instantons of the $(2,2)$ model with the rank deformation U$(N)_k\times$U$(N+M)_0\times$U$(N+2M)_{-k}\times$U$(N+M)_0$ are given by (see (3.20) in \cite{MNN})
\begin{align}
&\delta_1^{(2,2)}(k,M)=\frac{4\cos\frac{M\pi}{k}}{\sin^2\frac{\pi}{k}},\quad
\delta_2^{(2,2)}(k,M)=-\frac{4+\cos\frac{2M\pi}{k}}{\sin^2\frac{\pi}{k}},\quad
\delta_3^{(2,2)}(k,M)=\frac{12\cos\frac{M\pi}{k}}{\sin^2\frac{\pi}{k}},\nonumber\\
&\delta_4^{(2,2)}(k,M)=-\frac{32+16\cos\frac{2M\pi}{k}}{\sin^2\frac{\pi}{k}}+5,\quad
\delta_5^{(2,2)}(k,M)=\frac{220\cos\frac{M\pi}{k}+20\cos\frac{3M\pi}{k}}
{\sin^2\frac{\pi}{k}}-96\cos\frac{M\pi}{k}.
\label{22ws}
\end{align}
Comparing these functions with the worldsheet coefficients of the $(2,1)$ model \eqref{21ws1to7}, it is interesting to observe a close relation.
Namely, if we replace $k$ by $k/2$ and set $M=\pm 1/2$ in \eqref{22ws}, we can reproduce the worldsheet instantons of the $(2,1)$ model \eqref{21ws1to7} correctly
\begin{align}
\delta_d(k)=\delta^{(2,2)}_d\biggl(\frac{k}{2},\pm\frac{1}{2}\biggr).
\label{22&21}
\end{align}

This observation explains the match in lower instantons and the mismatch in higher instantons observed in \cite{MN3}.
The relation observed in \cite{MN3} is correctly reproduced in lower instantons if we assume the relation \eqref{22&21}.
Since the cosine functions in the numerator of \eqref{22ws} comes from the rank deformation, setting the cosine functions in $\delta_d(k)$ to be $1$ amounts to changing $M=\pm 1/2$ to $M=0$.
When we proceed to higher instantons and perform the replacement \eqref{22&21}, the numerator of \eqref{22ws} contains the cosine functions with larger arguments, which cause the mismatch after being reexpanded by the denominator $\sin^2\frac{\pi}{k}$.

There is an important implication from this observation.
Though in \cite{MN1} and \cite{MN3} it was difficult to see whether the non-perturbative part fits to \eqref{np}, with the expression of the K\"ahler parameters for the rank-deformed $(2,2)$ model \cite{MNN}
\begin{align}
T^\pm=\frac{\mu_\text{eff}}{k}\pm\pi i\biggl(1-\frac{M}{k}\biggr),
\end{align}
the relation \eqref{22&21} means that we can give a general expression for the worldsheet instanton if we choose the K\"ahler parameters and the string coupling constant schematically as
\begin{align}
T\sim\frac{2\mu_\text{eff}}{k}\pm\pi i\pm\frac{\pi i}{k},\quad
g_\text{s}\sim\frac{2}{k}.
\label{Tgs}
\end{align}
If we look at the membrane instanton more carefully, however, the fit to the expression \eqref{np} is not so trivial since the odd membrane instantons in \eqref{21mbexpand} does not have the linear term in $\mu_\text{eff}$.
A naive idea would be the cancellation between $e^{-\frac{T}{g_\text{s}}}\sim e^{-\mu_\text{eff}}e^{\pm\frac{\pi i}{2}}$, though a careful study shows that the cancellation does not work due to the extra factor in $e^{-\frac{T}{g_\text{s}}}\sim e^{-\mu_\text{eff}}e^{\pm\frac{\pi ki}{2}}e^{\pm\frac{\pi i}{2}}$.
This problem, in turn, can be solved by introducing all of the four K\"ahler parameters in \eqref{Tgs}.
In fact, with this setup, we shall see later in section \ref{21model4Kahler} that the cancellation happens beautifully.
The introduction of the four K\"ahler parameters is partially motivated by the study of the ${\mathbb Z}_2$ orbifold of the ABJM theory, or the $(1,1,1,1)$ model, in \cite{MNN}.
In relating this model to the $(2,2)$ model by changing the brane configuration, we found a non-trivial cancellation of odd instantons, which is very similar to the cancellation of the linear $\mu_\text{eff}$ term here.

\subsection{Multi-covering structure for membrane instantons}\label{mbmc}

Once we have found the relation to the $(2,2)$ model in the worldsheet instantons, we are motivated to relate the membrane instantons of the $(2,1)$ model with those of the $(2,2)$ model as well.
Interestingly, we find that the even membrane instantons \eqref{btildewithoutMC} possess the following novel multi-covering structure
\begin{align}
{\widetilde b}_{2\ell}
=\sum_{n|2\ell, n\in 2\mathbb{N}}\frac{(-1)^\ell}{n}\beta_{\frac{2\ell}{n}}(nk)
+\sum_{n|2\ell, n\in 2\mathbb{N}-1}\frac{1}{n}\beta_{\frac{2\ell}{n}}'(nk),
\label{bevenmc}
\end{align}
or explicitly
\begin{align}
&\widetilde b_2=-\frac{1}{2}\beta_1(2k)+\beta'_2(k),\quad
\widetilde b_4=\frac{1}{4}\beta_1(4k)+\frac{1}{2}\beta_2(2k)+\beta'_4(k),
\nonumber\\
&\widetilde b_6=-\frac{1}{6}\beta_1(6k)+\frac{1}{3}\beta'_2(3k)
-\frac{1}{2}\beta_3(2k)+\beta'_6(k),\quad
\cdots,
\label{beven}
\end{align}
where $\beta_d(k)$ is defined from the membrane instanton coefficient $\beta_d^{(2,2)}(k)$ of the $(2,2)$ model without rank deformations (see (3.15) and (3.17) in \cite{MN3}) as
\begin{align}
\beta_d(k)=\beta_d^{(2,2)}\biggl(\frac{k}{2}\biggr).
\label{b21fromb22}
\end{align}
Indeed, in these expansions the new component $\beta'_d(k)$ at each order takes the form of
\begin{align}
\beta'_d(k)=\frac{\sum_n\beta_{d,n}\sin\pi nk}{2\pi\sin^2\frac{\pi k}{2}},
\end{align}
with a finite number of positive integers $\beta_{d,n}$, as in the case of the ABJM theory and the $(2,2)$ model.
Once we adopt this new multi-covering structure, we can also determine the coefficients of even instantons $\beta'_{d}(k)$ of higher degrees $d$.
The explicit expressions of the functions $\beta'_d(k)$ are summarized in appendix \ref{21instanton}, where the expressions of $\beta_d(k)$ are also recapitulated.

The above novel multi-covering structure \eqref{bevenmc} can be understood from the pole cancellation.
As our goal is to express the instanton effects as the free energy of topological strings \eqref{np} where the pole cancellation occurs among the multi-covering components of each degree without mixing, it is reasonable to require the instanton coefficients to have the same substructure.
The multi-covering structure \eqref{bevenmc} assisted with $\beta_d(k)$, along with \eqref{wsmc} and \eqref{coddmc}, is very important to respect this substructure of the pole cancellation.
For example let us consider the multi-covering component of degree $d=2$ in the instanton coefficient of $e^{-4\mu_\text{eff}}$ at $k=2$.
If we adopted $\beta'_2(k)$ coming directly from $\widetilde b_2$ for the multi-covering component of ${\widetilde b}_4$ at degree $d=2$, the poles in the combination
\begin{align}
\frac{1}{2}\delta_2\biggl(\frac{k}{2}\biggr)e^{-\frac{8\mu_\text{eff}}{k}}
+\biggl(\mu_\text{eff}-k^2\frac{d}{dk}\frac{1}{4k}\biggr)
\frac{1}{2}\beta'_2(2k)e^{-4\mu_\text{eff}},
\end{align}
in the limit $k\rightarrow 2$ were not cancelled any more.
The reason of adopting the multi-covering structure \eqref{bevenmc} will be explained more carefully from the viewpoint of the free energy of topological strings \eqref{np} in section \ref{21model4Kahler}.

\subsection{Group-theoretical viewpoint}\label{grouptheory}

Before proceeding to the analysis, we shall explain another interesting observation.
In \cite{MNN} it was found that the total BPS indices identified in \cite{HKP} are split due to the introduction of two K\"ahler parameters.
We recapitulate the BPS indices discovered in \cite{MNN} in table \ref{BPS}, though the table is rearranged in a different way.
With this rearrangement it is not difficult to find the relation to the decomposition of the representations in the algebra so$(10)$ to the subalgebra so$(8)\times$u$(1)$.
For example, the spin $(0,\frac{3}{2})$ sector of degree $4$ is reminiscent of the decomposition of the adjoint representation ${\bf 45}$ and the spin $(0,2)$ sector of degree $5$ is the decomposition of the representation ${\bf 144}$
\begin{align}
{\bf 45}&\to({\bf 8_v})_{+2}+({\bf 28})_0+({\bf 1})_0+({\bf 8_v})_{-2},\nonumber\\
{\bf 144}&\to({\bf 8_{s/c}})_{+3}+({\bf 56_{s/c}})_{+1}+({\bf 8_{s/c}})_{+1}
+({\bf 56_{s/c}})_{-1}+({\bf 8_{s/c}})_{-1}+({\bf 8_{s/c}})_{-3}.
\label{45&144}
\end{align}
Hence, the BPS index $29$ in table \ref{BPS} should be interpreted as the representations ${\bf 28}$ and ${\bf 1}$, while $64$ is interpreted as the representations ${\bf 56_{s/c}}$ and ${\bf 8_{s/c}}$.

\begin{table}[ht!]
\begin{center}
\begin{tabular}{|c|c|c|c|c|}
\hline
$d$&$(j_\text{L},j_\text{R})$&\text{BPS}&$(-1)^{d-1}\sum_{|{\bm d}|=1}
\bigl(N^{\bm d}_{j_\text{L},j_\text{R}}\bigr)_{d^+-d^-}$&representations\\
\hline\hline
$1$&$(0;0)$&$16$&$8_{+1}+8_{-1}$
&${\bf 16}$\\
\hline
$2$&$(0,\frac{1}{2})$&$10$&$1_{+2}+8_0+1_{-2}$
&${\bf 10}$\\
\hline
$3$&$(0,1)$&$16$&$8_{+1}+8_{-1}$
&${\bf 16}$\\
\hline
$4$&$(0,\frac{1}{2})$&$1$&$1_0$
&${\bf 1}$\\
\cline{2-5}
&$(0,\frac{3}{2})$&$45$&$8_{+2}+29_0+8_{-2}$
&${\bf 45}$\\
\cline{2-5}
&$(\frac{1}{2},2)$&$1$&$1_0$
&${\bf 1}$\\
\hline
$5$&$(0,1)$&$16$&$8_{+1}+8_{-1}$
&${\bf 16}$\\
\cline{2-5}
&$(0,2)$&$144$&$8_{+3}+64_{+1}+64_{-1}+8_{-3}$
&${\bf 144}$\\
\cline{2-5}
&$(\frac{1}{2},\frac{5}{2})$&$16$&$8_{+1}+8_{-1}$
&${\bf 16}$\\
\hline
\end{tabular}
\caption{The BPS indices $N^{\bm d}_{j_\text{L},j_\text{R}}$ for $1\le d\le 5$ of the $(2,2)$ model with the rank deformation U$(N)_k\times$U$(N+M)_0\times$U$(N+2M)_{-k}\times$U$(N+M)_0$.
The information on the non-vanishing BPS indices in the first three columns is recapitulated from the tables in \cite{HKP} and the split into various degree differences in the fourth column comes from \cite{MNN}.}
\label{BPS}
\end{center}
\end{table}

Reversely, after assuming that the BPS indices are obtained by decomposing the so$(10)$ representations to the subalgebra so$(8)\times$u$(1)$, with table \ref{so10rep} of the decomposition of various irreducible so$(10)$ representations, we can check that no other candidate combinations of the so$(10)$ representations can form the BPS indices $45$ or $144$ with the same decomposition.
This is true also for the other BPS indices.
We have listed the representations in table \ref{BPS}.
Though in \cite{HKP} the representations seem determined directly from the Weyl orbits, our determination of the representations is rather indirect through the decomposition.

It is known that the lattice points in the weight lattice with the identification of the root lattice are classified by the congruency class ${\mathbb Z}_4$ for so$(10)$, so are the irreducible representations.
It is interesting to further observe that the representations of so$(10)$ appearing in the total degree $d$ are all the representations in the congruency class of $d$ mod $4$.
For example, the representations appearing for odd $d$ are all fermionic ones with the dimensions being multiples of $16$.
For this reason, from now on our tables of the decomposition of the so$(10)$ representations and the characters in appendix \ref{D5data} are listed by the congruency class.

This observation for the BPS indices of the $(2,2)$ model from the group-theoretical viewpoint may apply not just to the $(2,2)$ model.
We also expect the group-theoretical viewpoint to work later in our study of the $(2,1)$ model.

\section{Topological string}\label{top}

In this section we shall see that the instanton effects of the $(2,1)$ model are consistent with the free energy of topological strings \eqref{np}.
First we provide a set of four K\"{a}hler parameters which realizes the following structures of the instanton coefficients,
\begin{itemize}
\item
the multi-covering structures of $d_\ell$ \eqref{wsmc}, ${\widetilde b}_{2\ell}$ \eqref{bevenmc} and ${\widetilde c}_{2\ell-1}$ \eqref{coddmc},
\item
the vanishing odd coefficients, ${\widetilde b}_{2\ell-1}=0$, and
\item
the derivative relation between ${\widetilde c}_{2\ell}$ and ${\widetilde b}_{2\ell}$ \eqref{derivative}.
\end{itemize}
Then we determine the BPS indices for small degrees.
Interestingly, the BPS indices again correspond to the decomposition of the so$(10)$ representations, where two differences of the degrees specifying the split of the BPS indices are identified with the two u$(1)$ charges in the decomposition to the subalgebra so$(6)\times$u$(1)\times$u$(1)$.
This is how the observations in section \ref{obs} are brought to life.
Furthermore, once the representations are determined from the $(2,2)$ model, this enables us a top-down derivation for all of the instanton coefficients of the $(2,1)$ model.

\subsection{K\"ahler parameters}\label{21model4Kahler}

Our starting point is the same topological string free energy \eqref{np}
\begin{align}
J^\text{WS}(\mu_\text{eff})&=\sum_{j_\text{L},j_\text{R}}\sum_{\bm d}N^{\bm d}_{j_\text{L},j_\text{R}}
\sum_{n=1}^\infty\frac{(-1)^{(s_\text{L}+s_\text{R}-1)n}s_\text{R}\sin 2\pi g_\text{s}ns_\text{L}}
{n(2\sin\pi g_\text{s}n)^2\sin 2\pi g_\text{s}n}e^{-n{\bm d}\cdot{\bm T}},\nonumber\\
J^\text{MB}(\mu_\text{eff})&=\sum_{j_\text{L},j_\text{R}}\sum_{\bm d}N^{\bm d}_{j_\text{L},j_\text{R}}
\sum_{n=1}^\infty\frac{\partial}{\partial g_\text{s}}
\biggl[g_\text{s}\frac{-\sin\frac{\pi n}{g_\text{s}}s_\text{L}\sin\frac{\pi n}{g_\text{s}}s_\text{R}}
{4\pi n^2(\sin\frac{\pi n}{g_\text{s}})^3}e^{-n\frac{{\bm d}\cdot{\bm T}}{g_\text{s}}}\biggr].
\label{WSMB}
\end{align}
The main assumption is to introduce the following four K\"ahler parameters
\begin{align}
T_\uparrow^\pm=\frac{2\mu_\text{eff}}{k}\pm\pi i\biggl(1+\frac{1}{k}\biggr),\quad
T_\downarrow^\pm=\frac{2\mu_\text{eff}}{k}\pm\pi i\biggl(-1+\frac{1}{k}\biggr),
\label{4K}
\end{align}
with the string coupling constant identified as $g_\text{s}=2/k$.
Due to the relation
\begin{align}
{\bm d}\cdot{\bm T}=d\frac{2\mu_\text{eff}}{k}
+d_\text{m}\pi i+d_\text{w}\frac{\pi i}{k},
\end{align}
with
\begin{align}
d=\sum_\pm(d^\pm_\uparrow+d^\pm_\downarrow),\quad
d_\text{m}=(d^+_\uparrow-d^+_\downarrow)-(d^-_\uparrow-d^-_\downarrow),\quad
d_\text{w}=(d^+_\uparrow+d^+_\downarrow)-(d^-_\uparrow+d^-_\downarrow),
\end{align}
we find that the whole information on the degrees ${\bm d}$ is simply encoded\footnote{The names of membrane degrees and worldsheet degrees will be clear in the later discussion.
The even/odd parities of these degrees all coincide.} in the total degree $d$, the {\it membrane} degree $d_\text{m}$ and the {\it worldsheet} degree $d_\text{w}$.
Hence, hereafter we sum the BPS indices over all degrees giving the same set of $(d,d_\text{w},d_\text{m})$ and label the BPS indices by these degrees
\begin{align}
N^{(d,d_\text{w},d_\text{m})}_{j_\text{L},j_\text{R}}
=\sum_{\{{\bm d}|(d,d_\text{w},d_\text{m})\}}N^{\bm d}_{j_\text{L},j_\text{R}}.
\end{align}
For our later analysis we further assume the even property of $2j_\text{L}+2j_\text{R}-1-d$ and the symmetry of the BPS indices
\begin{align}
N_{j_\text{L},j_\text{R}}^{(d,d_\text{w},d_\text{m})}
&=N_{j_\text{L},j_\text{R}}^{(d,-d_\text{w},d_\text{m})}.
\label{symdw}
\end{align}

Let us deduce the instanton coefficients from \eqref{WSMB}.
As was noticed in \cite{MN5}, the imaginary part $\pm\pi i$ in the K\"{a}hler parameters \eqref{4K} realizes the multi-covering structure of the worldsheet instanton \eqref{wsmc} when $2j_\text{L}+2j_\text{R}-1-d$ is even,
\begin{align}
J^\text{WS}(\mu_\text{eff})=\sum_{m=1}^\infty d_m
e^{-m\frac{2\mu_\text{eff}}{k}},\quad
d_m=\sum_{n|m}
\frac{1}{n}\delta_{\frac{m}{n}}\biggl(\frac{k}{n}\biggr),
\end{align}
where the multi-covering component of the worldsheet instanton is described by the BPS indices summed over all of the membrane degrees
\begin{align}
\delta_d(k)=\sum_{d_\text{w}}\sum_{j_\text{L},j_\text{R}}
N^{(d,d_\text{w})}_{j_\text{L},j_\text{R}}
\biggl[\frac{s_\text{R}\sin\frac{4\pi}{k}s_\text{L}}
{(2\sin\frac{2\pi}{k})^2\sin\frac{4\pi}{k}}
e^{-d_\text{w}\frac{\pi i}{k}}\biggr],\quad
N^{(d,d_\text{w})}_{j_\text{L},j_\text{R}}
=\sum_{d_\text{m}}N^{(d,d_\text{w},d_\text{m})}_{j_\text{L},j_\text{R}}.
\label{deltaBPS}
\end{align}

The membrane instanton coefficients can be read off from \eqref{WSMB} as
\begin{align}
J^\text{MB}(\mu_\text{eff})&=\sum_{\ell=1}^\infty
({\widetilde b}_\ell\mu_\text{eff}+{\widetilde c}_\ell)e^{-\mu_\text{eff}},
\end{align}
with ${\widetilde b}_\ell$ and ${\widetilde c}_\ell$ given respectively by
\begin{align}
{\widetilde b}_\ell
&=\sum_{nd=\ell}\sum_{j_\text{L},j_\text{R}}\sum_{d_\text{w}}\sum_{d_\text{m}}
N^{(d,d_\text{w},d_\text{m})}_{j_\text{L},j_\text{R}}
e^{-nd_\text{m}\frac{\pi ki}{2}}e^{-nd_\text{w}\frac{\pi i}{2}}
\frac{-d\sin\frac{\pi kn}{2}s_\text{L}\sin\frac{\pi kn}{2}s_\text{R}}
{4\pi n(\sin\frac{\pi kn}{2})^3},
\label{btilde}
\end{align}
and
\begin{align}
{\widetilde c}_\ell&=\sum_{nd=\ell}\sum_{j_\text{L},j_\text{R}}
\sum_{d_\text{w}}\sum_{d_\text{m}}
N^{(d,d_\text{w},d_\text{m})}_{j_\text{L},j_\text{R}}
e^{-nd_\text{m}\frac{\pi ki}{2}}e^{-nd_\text{w}\frac{\pi i}{2}}
\nonumber\\
&\qquad\times
\biggl(\frac{\pi i(kd_\text{m}+d_\text{w})}{2}-k^2\frac{d}{dk}\frac{1}{kn}\biggr)
\biggl[
\frac{-\sin\frac{\pi kn}{2}s_\text{L}\sin\frac{\pi kn}{2}s_\text{R}}
{4\pi n(\sin\frac{\pi kn}{2})^3}
\biggr].
\label{MBsubstitute}
\end{align}

Now we can see the vanishing of odd coefficients ${\widetilde b}_{2\ell-1}=0$ is realized from the symmetry of the BPS indices \eqref{symdw}.
This symmetry allows us to replace $e^{-nd_\text{w}\frac{\pi i}{2}}$ in \eqref{btilde} with $(e^{-nd_\text{w}\frac{\pi i}{2}}+e^{nd_\text{w}\frac{\pi i}{2}})/2$, which vanishes when $nd=2\ell-1$ is odd since $n$, $d$, $d_\text{w}$ are all odd.
Moreover, for ${\widetilde b}_{2\ell}$, by noticing
\begin{align}
\frac{e^{-nd_\text{w}\frac{\pi i}{2}}+e^{nd_\text{w}\frac{\pi i}{2}}}{2}
\Bigr|_{nd=2\ell}=\begin{cases}
(-1)^\ell,&\text{for even }n,\\
(-1)^{\frac{d_\text{w}}{2}},&\text{for odd }n,
\end{cases}
\end{align}
we obtain the following multi-covering structure
\begin{align}
{\widetilde b}_{2\ell}
=\sum_{n|2\ell,n\in 2\mathbb{N}}\frac{(-1)^\ell}{n}\beta_{\frac{2\ell}{n}}(nk)
+\sum_{n|2\ell,n\in 2\mathbb{N}-1}\frac{1}{n}\beta'_{\frac{2\ell}{n}}(nk),
\end{align}
which is exactly what we have suggested in \eqref{bevenmc}.
Here the multi-covering components are
\begin{align}
\beta_d(k)&=\sum_{j_\text{L},j_\text{R}}\sum_{d_\text{m}}N^{(d,d_\text{m})}_{j_\text{L},j_\text{R}}
\frac{-d\sin\frac{\pi k}{2}s_\text{L}\sin\frac{\pi k}{2}s_\text{R}}{4\pi(\sin\frac{\pi k}{2})^3}
e^{-d_\text{m}\frac{\pi ki}{2}},\nonumber\\
\beta'_d(k)&=\sum_{j_\text{L},j_\text{R}}\sum_{d_\text{m}}N'^{(d,d_\text{m})}_{j_\text{L},j_\text{R}}
\frac{-d\sin\frac{\pi k}{2}s_\text{L}\sin\frac{\pi k}{2}s_\text{R}}{4\pi(\sin\frac{\pi k}{2})^3}
e^{-d_\text{m}\frac{\pi ki}{2}},
\label{betas}
\end{align}
with the original BPS indices $N_{j_\text{L},j_\text{R}}^{(d,d_\text{m})}$ and the alternating BPS indices $N_{j_\text{L},j_\text{R}}'^{(d,d_\text{m})}$ defined as
\begin{align}
N^{(d,d_\text{m})}_{j_\text{L},j_\text{R}}
=\sum_{d_\text{w}}
N^{(d,d_\text{w},d_\text{m})}_{j_\text{L},j_\text{R}},\quad
N'^{(d,d_\text{m})}_{j_\text{L},j_\text{R}}
=\Biggl(\sum_{d_\text{w}\equiv 0\,({\rm mod}4)}
-\sum_{d_\text{w}\equiv 2\,({\rm mod}4)}\Biggr)
N^{(d,d_\text{w},d_\text{m})}_{j_\text{L},j_\text{R}}.
\label{alternating}
\end{align}

The coefficient ${\widetilde c}_\ell$ \eqref{MBsubstitute} can be simplified in the same way.
For even instantons ${\widetilde c}_{2\ell}$, from the symmetry of the BPS indices \eqref{symdw}, the $\pi id_\text{w}/2$ term is cancelled and the $\pi ikd_\text{m}/2$ term is combined into the derivative term to reproduce the derivative relation \eqref{derivative}.
For odd instantons ${\widetilde c}_{2\ell-1}$, on the other hand, from the symmetry of the BPS indices \eqref{symdw}, only the $\pi id_\text{w}/2$ term  survives.
Due to the simplification
\begin{align}
\frac{id_\text{w}e^{-nd_\text{w}\frac{\pi i}{2}}
-id_\text{w}e^{nd_\text{w}\frac{\pi i}{2}}}{2}\biggr|_{nd=2\ell-1}
=(-1)^{\frac{n-1}{2}+\frac{d_\text{w}-1}{2}}d_\text{w},
\end{align}
(which can be proved from $(e^{\pm\frac{\pi i}{2}})^{(n-1)(d_\text{w}-1)}=1$ by noting both $n$ and $d_\text{w}$ are odd if $nd$ is odd), we reproduce the multi-covering structure \eqref{coddmc}
\begin{align}
{\widetilde c}_{2\ell-1}=\sum_{n|2\ell-1}\frac{(-1)^{\frac{n-1}{2}}}{n}\gamma_{\frac{2\ell-1}{n}}(nk),
\end{align}
where the multi-covering components are
\begin{align}
\gamma_d(k)=\sum_{j_\text{L},j_\text{R}}\sum_{d_\text{m}}M_{j_\text{L},j_\text{R}}^{(d,d_\text{m})}\frac{-\sin\frac{\pi k}{2}s_\text{L}\sin\frac{\pi k}{2}s_\text{R}}{8(\sin\frac{\pi k}{2})^3}e^{-d_\text{m}\frac{\pi ki}{2}},
\label{gammas}
\end{align}
with the weighted BPS indices $M_{j_\text{L},j_\text{R}}^{(d,d_\text{m})}$
\begin{align}
M^{(d,d_\text{m})}_{j_\text{L},j_\text{R}}
=\sum_{d_\text{w}}(-1)^{\frac{d_\text{w}-1}{2}}d_\text{w}
N^{(d,d_\text{w},d_\text{m})}_{j_\text{L},j_\text{R}}.
\label{weighted}
\end{align}

\subsection{BPS indices}\label{BPSindex}

After constructing the general framework to reproduce the multi-covering structure and the derivative relation, now we can ask whether the expression of the topological string free energy matches with the instanton coefficients if we choose the BPS indices suitably.
As in \cite{MNN} we shall assume the positivity $(-1)^{d-1}N^{(d,d_\text{w},d_\text{m})}_{j_\text{L},j_\text{R}}\ge 0$ and study how the original total BPS indices listed in \cite{HKP} is partitioned
\begin{align}
N^d_{j_\text{L},j_\text{R}}=\sum_{d_\text{w}}\sum_{d_\text{m}} N^{(d,d_\text{w},d_\text{m})}_{j_\text{L},j_\text{R}}.
\end{align}

We have observed in \eqref{22&21} that the worldsheet instantons of the rank deformed $(2,2)$ model agree with those of the $(2,1)$ model if we rescale $k$ by $1/2$ and set $M=\pm 1/2$.
Hence, if the worldsheet BPS indices $N^{(d,d_\text{w})}_{j_\text{L},j_\text{R}}$ are those identified in table \ref{BPS}, this expression automatically reproduces the worldsheet instantons of the $(2,1)$ model.
Also, since we have brought the expression of $\beta_d(k)$ from the $(2,2)$ model as in \eqref{b21fromb22}, we also hope to identify the membrane BPS indices $N^{(d,d_\text{m})}_{j_\text{L},j_\text{R}}$ to be those in table \ref{BPS}.

For $d=1,2,3$, since there is only one type of spins for each degree we find the identification
\begin{align}
M^{(1,\pm 1)}_{0,0}=4,\qquad
N'^{(2,\pm 2)}_{0,\frac{1}{2}}=-1,\quad N'^{(2,0)}_{0,\frac{1}{2}}=-4,\qquad
M^{(3,\pm 1)}_{0,1}=4,
\end{align}
from the comparison of the general expression \eqref{betas}, \eqref{gammas} with $\gamma_1(k)$, $\beta'_2(k)$, $\gamma_3(k)$.
Combining with the condition of the total worldsheet BPS indices $N^{(d,d_\text{w})}_{j_\text{L},j_\text{R}}$ and the total membrane BPS indices $N^{(d,d_\text{m})}_{j_\text{L},j_\text{R}}$, both of which are given by
\begin{align}
&N^{(1,d_\text{w}=\pm 1)}_{0,0}=8,\qquad
N^{(2,d_\text{w}=\pm 2)}_{0,\frac{1}{2}}=-1,\quad
N^{(2,d_\text{w}=0)}_{0,\frac{1}{2}}=-8,\qquad
N^{(3,d_\text{w}=\pm 1)}_{0,1}=8,\nonumber\\
&N^{(1,d_\text{m}=\pm 1)}_{0,0}=8,\qquad
N^{(2,d_\text{m}=\pm 2)}_{0,\frac{1}{2}}=-1,\quad
N^{(2,d_\text{m}=0)}_{0,\frac{1}{2}}=-8,\qquad
N^{(3,d_\text{m}=\pm 1)}_{0,1}=8,
\end{align}
we find that the separated BPS indices $N^{(d,d_\text{w},d_\text{m})}_{j_\text{L},j_\text{R}}$ are
\begin{align}
N^{(1,\pm 1,\pm 1)}_{0,0}=4,\qquad
N^{(2,\pm 2,0)}_{0,\frac{1}{2}}=N^{(2,0,\pm 2)}_{0,\frac{1}{2}}=-1,\quad
N^{(2,0,0)}_{0,\frac{1}{2}}=-6,\qquad
N^{(3,\pm 1,\pm 1)}_{0,1}=4.
\end{align}
Looking closely at the decomposition for $d=2$, for example, we find that the membrane BPS index $|N^{(2,d_\text{m}=0)}_{0,\frac{1}{2}}|=8$ is split into
\begin{align}
8\to 1_{+2}+6_0+1_{-2},
\end{align}
where we have denoted the worldsheet degree $d_\text{w}$ of $N^{(d,d_\text{w},d_\text{m})}_{j_\text{L},j_\text{R}}$ as subscripts.
Then, this expression is reminiscent of the decomposition of the representation ${\bf 8_v}$ from so$(8)$ to the subalgebra so$(6)\times$u$(1)$.
This interpretation works for the other BPS indices in $d=1,2,3$ as well.

After observing the relation to the further decomposition of the so$(8)$ representations to so$(6)\times$u$(1)$, since we have already identified the BPS indices as the representations of so$(10)$ for $d=4,5$ in table \ref{BPS}, the only remaining task is to decompose each so$(8)$ representation in \eqref{45&144} to so$(6)\times$u$(1)$,
\begin{align}
&{\bf 28}\to({\bf 6})_{+2}+({\bf 15})_0+({\bf 1})_0+({\bf 6})_{-2},\quad
{\bf 8_v}\to({\bf 1})_{+2}+({\bf 6})_0+({\bf 1})_{-2},\quad
{\bf 1}\to({\bf 1})_0,\nonumber\\
&{\bf 56_{s/c}}\to({\bf 4})_{+3}+(\overline{\bf 20})_{+1}+(\overline{\bf 4})_{+1}
+({\bf 20})_{-1}+({\bf 4})_{-1}+(\overline{\bf 4})_{-3},\quad
{\bf 8_{s/c}}\to(\overline{\bf 4})_{+1}+({\bf 4})_{-1}.
\end{align}
Then we find that the degrees should be decomposed as
\begin{align}
&N^{(4,0,0)}_{0,\frac{3}{2}}=-17,\quad
N^{(4,0,\pm 2)}_{0,\frac{3}{2}}=N^{(4,\pm 2,0)}_{0,\frac{3}{2}}=-6,\quad
N^{(4,\pm 2,\pm 2)}_{0,\frac{3}{2}}=-1,
\nonumber\\
&N^{(5,\pm 3,\pm 1)}_{0,2}=N^{(5,\pm 1,\pm 3)}_{0,2}=28,\quad
N^{(5,\pm 3,\pm 3)}_{0,2}=4,
\end{align}
which gives the alternating BPS indices and the weighted BPS indices
\begin{align}
&N'^{(4,0)}_{0,\frac{3}{2}}=-5,\quad
N'^{(4,\pm 2)}_{0,\frac{3}{2}}=-4,
\nonumber\\
&M^{(5,\pm 1)}_{0,2}=32,\quad
M^{(5,\pm 3)}_{0,2}=8.
\end{align}
Substituting these BPS indices into \eqref{betas} and \eqref{gammas}, we find that the instanton coefficients in \eqref{21mbeven} and \eqref{21mbodd} obtained from the WKB expansions are beautifully reproduced.

To summarize, our proposal is that the reduced grand potential of the $(2,1)$ model is described by the BPS indices which are obtained by identifying the total BPS indices of the local $D_5$ del Pezzo geometry as the representations of so$(10)$ and decomposing the so$(10)$ representations to the subalgebra so$(6)\times$u$(1)\times$u$(1)$ with the two u$(1)$ charges identified as the two degree differences.

\subsection{Characters}

We have found that we can describe the reduced grand potential of the $(2,1)$ model by the free energy of topological strings if we adopt the ansatz of the four K\"ahler parameters \eqref{4K} and choose the BPS indices appropriately by the decomposition of the so$(10)$ representations.
Here we point out that our proposal on the reduced grand potential can be summarized compactly in terms of the characters of so$(10)$.

For this purpose, we first introduce the characters of so$(10)$ with two fugacities,
\begin{align}
\chi_{\bf R}(p,q)=\sum_{d_\text{w},d_\text{m}}
p^{d_\text{w}}q^{d_\text{m}}\dim\bigl({\bf r}_{(d_\text{w},d_\text{m})}\bigr),
\end{align}
each of which measures the two u$(1)$ charges in the decomposition
\begin{align}
\text{so}(10)\to\text{so}(6)\times\text{u}(1)\times\text{u}(1),\quad
{\bf R}\to\sum {\bf r}_{(d_\text{w},d_\text{m})}.
\end{align}
Then, once the total BPS index is identified as the so$(10)$ representations,
\begin{align}
(-1)^{d-1}N^d_{j_\text{L},j_\text{R}}=\sum_{\bf R}n^{d,{\bf R}}_{j_\text{L},j_\text{R}}
\dim({\bf R}),
\end{align}
each BPS index coming from the so$(10)$ representations can be given as
\begin{align}
(-1)^{d-1}\sum_{d_\text{w}}N^{(d,d_\text{w})}_{j_\text{L},j_\text{R}}p^{d_\text{w}}
&=\sum_{\bf R}n^{d,{\bf R}}_{j_\text{L},j_\text{R}}
\chi_{\bf R}(p,1),&
(-1)^{d-1}\sum_{d_\text{m}}N^{(d,d_\text{m})}_{j_\text{L},j_\text{R}}q^{d_\text{m}}
&=\sum_{\bf R}n^{d,{\bf R}}_{j_\text{L},j_\text{R}}
\chi_{\bf R}(1,q),\nonumber\\
(-1)^{d-1}\sum_{d_\text{m}}N'^{(d,d_\text{m})}_{j_\text{L},j_\text{R}}q^{d_\text{m}}
&=\sum_{\bf R}n^{d,{\bf R}}_{j_\text{L},j_\text{R}}
\chi_{\bf R}(i,q),&
(-1)^{d-1}\sum_{d_\text{m}}M^{(d,d_\text{m})}_{j_\text{L},j_\text{R}}q^{d_\text{m}}
&=\sum_{\bf R}n^{d,{\bf R}}_{j_\text{L},j_\text{R}}
\frac{\partial\chi_{\bf R}}{\partial p}(i,q).
\label{BPSch}
\end{align}
This implies from \eqref{deltaBPS}, \eqref{betas}, \eqref{gammas} that the multi-covering components of the worldsheet instantons and the membrane instantons are compactly given in terms of the characters by
\begin{align}
\delta_d(k)
&=\frac{(-1)^{d-1}}{(2\sin\frac{2\pi}{k})^2}\sum_{j_\text{L},j_\text{R}}
\sum_{\bf R}n^{d,{\bf R}}_{j_\text{L},j_\text{R}}
\chi_{\bf R}(e^{-\frac{\pi i}{k}},1)
\chi_{j_\text{L}}(e^{\frac{4\pi i}{k}})\chi_{j_\text{R}}(1),\nonumber\\
\beta_d(k)
&=\frac{(-1)^dd}{4\pi\sin\frac{\pi k}{2}}
\sum_{j_\text{L},j_\text{R}}\sum_{\bf R}n^{d,{\bf R}}_{j_\text{L},j_\text{R}}
\chi_{\bf R}(1,e^{-\frac{\pi ki}{2}})
\chi_{j_\text{L}}(e^{\frac{\pi ki}{2}})
\chi_{j_\text{R}}(e^{\frac{\pi ki}{2}}),\nonumber\\
\beta'_d(k)
&=\frac{(-1)^dd}{4\pi\sin\frac{\pi k}{2}}
\sum_{j_\text{L},j_\text{R}}\sum_{\bf R}n^{d,{\bf R}}_{j_\text{L},j_\text{R}}
\chi_{\bf R}(i,e^{-\frac{\pi ki}{2}})
\chi_{j_\text{L}}(e^{\frac{\pi ki}{2}})
\chi_{j_\text{R}}(e^{\frac{\pi ki}{2}}),\nonumber\\
\gamma_d(k)
&=
\frac{(-1)^d}{8\sin\frac{\pi k}{2}}
\sum_{j_\text{L},j_\text{R}}\sum_{\bf R}n^{d,{\bf R}}_{j_\text{L},j_\text{R}}
\frac{\partial\chi_{\bf R}}{\partial p}(i,e^{-\frac{\pi ki}{2}})
\chi_{j_\text{L}}(e^{\frac{\pi ki}{2}})
\chi_{j_\text{R}}(e^{\frac{\pi ki}{2}}),
\end{align}
where we have also introduced the su$(2)$ character
\begin{align}
\chi_j(q)=\frac{q^{2j+1}-q^{-(2j+1)}}{q-q^{-1}}.
\end{align}

\subsection{Higher degrees}

We believe that all the evidences we have provided in section \ref{BPSindex} are already quite non-trivial.
Nevertheless, in this subsection we shall proceed to even higher degrees $d=6,7,8$ to convince the readers completely of our proposal.

\begin{table}[ht!]
\begin{center}
\begin{tabular}{|c|c|c|c|}
\hline
$d$&$(j_\text{L},j_\text{R})$&\text{BPS}&representations\\
\hline\hline
$6$&$(0,\frac{1}{2})$&$10$&${\bf 10}$\\
\cline{2-4}
&$(0,\frac{3}{2})$&$130$&${\bf 120}+{\bf 10}$\\
\cline{2-4}
&$(0,\frac{5}{2})$&$456$&${\bf 320}+{\bf 126}+{\bf 10}$\\
\cline{2-4}
&$(\frac{1}{2},2)$&$10$&${\bf 10}$\\
\cline{2-4}
&$(\frac{1}{2},3)$&$130$&${\bf 120}+{\bf 10}$\\
\cline{2-4}
&$(1,\frac{7}{2})$&$10$&${\bf 10}$\\
\hline
\end{tabular}\quad
\begin{tabular}{|c|c|c|c|}
\hline
$d$&$(j_\text{L},j_\text{R})$&\text{BPS}&representations\\
\hline\hline
$7$&$(0,0)$&$16$&${\bf 16}$\\
\cline{2-4}
&$(0,1)$&$160$&${\bf 144}+{\bf 16}$\\
\cline{2-4}
&$(0,2)$&$736$&${\bf 560}+{\bf 144}+2\times{\bf 16}$\\
\cline{2-4}
&$(0,3)$&$1440$&${\bf 720}+{\bf 560}+{\bf 144}+{\bf 16}$\\
\cline{2-4}
&$(0,4)$&$16$&${\bf 16}$\\
\cline{2-4}
&$(\frac{1}{2},\frac{3}{2})$&$16$&${\bf 16}$\\
\cline{2-4}
&$(\frac{1}{2},\frac{5}{2})$&$176$&${\bf 144}+2\times{\bf 16}$\\
\cline{2-4}
&$(\frac{1}{2},\frac{7}{2})$&$736$&${\bf 560}+{\bf 144}+2\times{\bf 16}$\\
\cline{2-4}
&$(1,3)$&$16$&${\bf 16}$\\
\cline{2-4}
&$(1,4)$&$160$&${\bf 144}+{\bf 16}$\\
\cline{2-4}
&$(\frac{3}{2},\frac{9}{2})$&$16$&${\bf 16}$\\
\hline
\end{tabular}
\caption{The constituent representations for the total BPS indices of the $(2,2)$ model for $d=6,7$.}
\label{BPS67}
\end{center}
\end{table}

After proposing to obtain the BPS indices from the decomposition of the representations, our remaining task is to identify the so$(10)$ representations which the total BPS indices listed in \cite{HKP} consist of and to decompose the representations to the subalgebra so$(8)\times$u$(1)$.
This can be done completely in the study of the $(2,2)$ model before considering the $(2,1)$ model.
Then, we can apply our rule of further decomposing the so$(8)$ representations to the subalgebra so$(6)\times$u$(1)$ to see whether the predicted worldsheet instantons coincide with those of the $(2,1)$ model obtained from the numerical fitting in \eqref{21ws1to7} and whether the predicted membrane instantons coincide with those of the $(2,1)$ model obtained from the WKB expansion in \eqref{21mbodd} and \eqref{21mbeven}.
Hence we start our analysis purely on the $(2,2)$ model. 

For $d=6$ we can study either from the numerical values of the worldsheet instantons of the $(2,2)$ model or the WKB expansion for the membrane instantons.
In either method, we assume that the total BPS index $456$ in the spin $(0,\frac{5}{2})$ can be given by an integral linear combination of all the representations in the congruency class of $6\equiv 2$ mod $4$ with the dimensions smaller than or equal to $456$ (which are ${\bf 10}$, ${\bf 120}$, ${\bf 126}$, ${\bf 210'}$ and ${\bf 320}$), while the total BPS indices $130$ in the spins $(0,\frac{3}{2})$ and $(\frac{1}{2},3)$ are given by other linear combinations of ${\bf 10}$, ${\bf 120}$ and ${\bf 126}$.
Then, for the former method, we ask which combination gives correctly the numerical values listed in appendix C.1.3 of \cite{MNN}, while for the latter method, we ask which combination gives correctly the WKB expansion in \eqref{beta22WKB}.
In either method, we obtain the result in table \ref{BPS67}.
For $d=7$ we need to utilize both the numerical values of the worldsheet instantons in appendix C.1.3 of \cite{MNN} and the WKB expansion of the membrane instantons in \eqref{beta22WKB}. 
With both the data we can again fix exactly which representations appear in the total BPS indices.
The results are listed in table \ref{BPS67}.

For $d=6,7$ we can substitute the BPS indices into the worldsheet instanton to find
\begin{align}
\delta_6(k,M)&=-\frac{756+579\cos\frac{2M\pi}{k}+24\cos\frac{4M\pi}{k}}
{\sin^2\frac{\pi}{k}}+\biggl(800+480\cos\frac{2M\pi}{k}\biggr)\nonumber\\
&\qquad-\biggl(256+64\cos\frac{2M\pi}{k}\biggr)\sin^2\frac{\pi}{k},\\
\delta_7(k,M)
&=\frac{7112\cos\frac{M\pi}{k}+1288\cos\frac{3M\pi}{k}+28\cos\frac{5M\pi}{k}}
{\sin^2\frac{\pi}{k}}
-\biggl(13120\cos\frac{M\pi}{k}+1696\cos\frac{3M\pi}{k}\biggr)\nonumber\\
&\qquad
+\biggl(9472\cos\frac{M\pi}{k}+576\cos\frac{3M\pi}{k}\biggr)\sin^2\frac{\pi}{k}
-2560\cos\frac{M\pi}{k}\sin^4\frac{\pi}{k}.
\end{align}
We find that we can obtain the worldsheet instanton of the $(2,1)$ model \eqref{21ws1to7} by substituting $M=\pm 1/2$ and replacing $k$ by $k/2$ as in \eqref{22&21}.
By applying this rule we encounter the cosine functions with higher arguments which can be reexpanded by the denominator, as we have explained below \eqref{22&21}.
Due to this reason, the rule observed in \cite{MN3} should be modified by \eqref{22&21}.

Now with the characters in appendix \ref{aw} which computes the alternating BPS indices $N'^{(d,d_\text{m})}_{j_\text{L},j_\text{R}}$ and the weighted BPS indices $M^{(d,d_\text{m})}_{j_\text{L},j_\text{R}}$ for various so$(10)$ representations, we can predict the membrane instantons of the $(2,1)$ model for $d=6,7$.
We find a very non-trivial match with those of the $(2,1)$ model \eqref{21mbodd} and \eqref{21mbeven} obtained by the WKB expansion.

\begin{table}[ht!]
\begin{center}
\begin{tabular}{|c|c|c|c|c|}
\hline
$d$&$(j_\text{L},j_\text{R})$&BPS&representations\\
\hline\hline
$8$&$(0,\frac{7}{2})$&$4726$
&${\bf 1386}+{\bf 1050}+2\times{\bf 945}+{\bf 210}+{\bf 54}
+3\times{\bf 45}+{\bf 1}$\\
\cline{2-4}
&$(0,\frac{5}{2}),(\frac{1}{2},4)$&$3431$
&${\bf 1050}+{\bf 945}+{\bf 770}+2\times{\bf 210}
+2\times{\bf 54}+3\times{\bf 45}+3\times{\bf 1}$\\
\cline{2-4}
&$(\frac{1}{2},3)$&$1602$
&${\bf 945}+2\times{\bf 210}+{\bf 54}
+4\times{\bf 45}+3\times{\bf 1}$\\
\cline{2-4}
&$(0,\frac{3}{2}),(1,\frac{9}{2})$&$1345$
&${\bf 945}+{\bf 210}+{\bf 54}+3\times{\bf 45}+{\bf 1}$\\
\cline{2-4}
&$(\frac{1}{2},2),(1,\frac{7}{2})$&$357$
&${\bf 210}+{\bf 54}+2\times{\bf 45}+3\times{\bf 1}$\\
\cline{2-4}
&$(0,\frac{1}{2}),(\frac{3}{2},5)$&$311$
&${\bf 210}+{\bf 54}+{\bf 45}+2\times{\bf 1}$\\
\cline{2-4}
&$(0,\frac{9}{2})$&$257$
&${\bf 210}+{\bf 45}+2\times{\bf 1}$\\
\cline{2-4}
&$\begin{array}{c}(\frac{1}{2},1),(\frac{1}{2},5),\\[-3pt]
(1,\frac{5}{2}),(\frac{3}{2},4)\end{array}$&$46$
&${\bf 45}+{\bf 1}$\\
\cline{2-4}
&$(2,\frac{11}{2})$&$45$&${\bf 45}$\\
\cline{2-4}
&$\begin{array}{c}(1,\frac{3}{2}),(1,\frac{11}{2}),(\frac{3}{2},3),\\[-3pt]
(2,\frac{9}{2}),(\frac{5}{2},6)\end{array}$&$1$
&${\bf 1}$\\
\hline
\end{tabular}
\caption{The constituent representations for the total BPS indices of the $(2,2)$ model for $d=8$.}
\label{BPS8}
\end{center}
\end{table}

For $d=8$, since there are more degrees of freedom to identify the representations, we need to impose one more assumption.
In the table of \cite{HKP} the total BPS indices $3431$ appear in both the spins $(0,\frac{5}{2})$ and $(\frac{1}{2},4)$.
We assume that the same numbers of the BPS indices in different spins are identified as the same combination of the so$(10)$ representations.
Under this assumption, we find only two solutions.
Aside from the one listed in table \ref{BPS8}, the other solution is to replace the representations for the total BPS indices $3431$ by
\begin{align}
2\times{\bf 1050}+{\bf 945}+2\times{\bf 54}+3\times{\bf 45}+143\times{\bf 1}.
\end{align}
From the characters in \eqref{chiso10cong0}, we find that only the set of representations listed in table \ref{BPS8} correctly reproduces the membrane instanton coefficient of the $(2,1)$ model \eqref{21mbeven} obtained by the WKB expansion.

In the above identification of the representations for the BPS indices of $d=8$, we have adopted the assumption that the same BPS indices consist of the same set of the so$(10)$ representations.
Since we do not have a persuasive reason for this assumption, we have also performed an alternative analysis.
Namely, instead of the above assumption, we adopt our proposal of the relation between the $(2,2)$ model and the decomposition of the so$(10)$ representations to so$(8)\times$u$(1)$ and the relation between the $(2,1)$ model and the decomposition of the same representations to so$(6)\times$u$(1)\times$u$(1)$ simultaneously.
Then, we reach the same result of the identification of the so$(10)$ representation listed in table \ref{BPS8}.

\section{Rank-deformed $(2,2)$ model from characters}\label{22}

Previously in \cite{MNN} two types of rank deformations in the $(2,2)$ model were studied.
As we have seen in section \ref{grouptheory}, one of the rank deformations U$(N)_k\times$U$(N+M)_0\times$U$(N+2M)_{-k}\times$U$(N+M)_0$ corresponds to introducing the fugacity to distinguish the u$(1)$ charge in the decomposition of the so$(10)$ representations to the subalgebra so$(8)\times$u$(1)$.
Here let us turn to revisiting the two-parameter rank deformation U$(N+M_\text{II})_k\times$U$(N+M_\text{I})_0\times$U$(N+2M_\text{I}+M_\text{II})_{-k}\times$U$(N+M_\text{I})_0$ in \cite{MNN} where the previous deformation corresponds to $(M_\text{I},M_\text{II})=(M,0)$.

To describe this deformation, in \cite{MNN} six K\"ahler parameters were identified
\begin{align}
T_1^\pm&=\frac{\mu_\text{eff}}{k}
\pm\pi i\biggl(1-\frac{M_\text{I}}{k}-\frac{2M_\text{II}}{k}\biggr),\nonumber\\
T_2^\pm&=\frac{\mu_\text{eff}}{k}\pm\pi i\biggl(1-\frac{M_\text{I}}{k}\biggr),
\nonumber\\
T_3^\pm&=\frac{\mu_\text{eff}}{k}
\pm\pi i\biggl(1-\frac{M_\text{I}}{k}+\frac{2M_\text{II}}{k}\biggr),
\end{align}
and the corresponding BPS indices were studied.
It was difficult to distribute the BPS indices into various degrees precisely, which is essentially due to the relations among the K\"ahler parameters
\begin{align}
2T_2^\pm=T_1^\pm+T_3^\pm,\quad
T_1^++T_1^-=T_2^++T_2^-=T_3^++T_3^-,\quad
T_2^\pm+T_1^\mp=T_2^\mp+T_3^\pm.
\end{align}
In other words, the description in \cite{MNN} with the six K\"ahler parameters is probably correct though it may not be the most economical description because the six K\"ahler parameters are too abundant for the deformation with only two parameters.
Our studies in the previous section suggest that instead of introducing many K\"ahler parameters it is more economical to identify the u$(1)$ charge correctly and describe the reduced grand potential by the characters with the u$(1)$ fugacity.
From this viewpoint, in addition to the previous u$(1)$ charge $d_\text{I}$ appearing in decomposing the so$(10)$ representations to so$(8)\times$u$(1)$, we introduce another u$(1)$ charge $d_\text{II}$, both of which are given explicitly in the current degrees by 
\begin{align}
d_\text{I}=(d_1^++d_2^++d_3^+)-(d_1^-+d_2^-+d_3^-),\quad
d_\text{II}=(d_1^+-d_1^-)-(d_3^+-d_3^-).
\end{align} 
With these two u$(1)$ charges we can rearrange table 2 and table 3 in \cite{MNN} by table \ref{anotherdeform} .

\begin{table}[ht!]
\begin{center}
\begin{tabular}{|c|c|c|c|c|}
\hline
$d$&$(j_\text{L},j_\text{R})$&$d_\text{I}$&\text{BPS}
&$(-1)^{d-1}\sum_{d_\text{II}}
\bigl(N^{(d,d_\text{I},d_\text{II})}_{j_\text{L},j_\text{R}}\bigr)_{d_\text{II}}$\\
\hline\hline
$1$&$(0;0)$&$\pm 1$&$8$&$2_{+1}+4_0+2_{-1}$\\
\hline
$2$&$(0,\frac{1}{2})$&$0$&$8$&$2_{+1}+4_0+2_{-1}$\\
\cline{3-5}
&&$\pm 2$&1&$1_0$\\
\hline
$3$&$(0,1)$&$\pm 1$&$8$&$2_{+1}+4_0+2_{-1}$\\
\hline
$4$&$(0,\frac{1}{2})$&$0$&$1$&$1_0$\\
\cline{2-5}
&$(0,\frac{3}{2})$&$0$&$29$&$1_{+2}+8_{+1}+11_0+8_{-1}+1_{-2}$\\
\cline{3-5}
&&$\pm 2$&$8$&$2_{+1}+4_0+2_{-1}$\\
\cline{2-5}
&$(\frac{1}{2},2)$&$0$&$1$&$1_0$\\
\hline
\end{tabular}
\caption{The BPS indices $N^{\bm d}_{j_\text{L},j_\text{R}}$ for $1\le d\le 4$ of the $(2,2)$ model with the rank deformation U$(N+M_\text{II})_k\times$U$(N+M_\text{I})_0\times$U$(N+2M_\text{I}+M_\text{II})_{-k}\times$U$(N+M_\text{I})_0$.
The table is recapitulated from the tables in \cite{MNN} with a different arrangement.}
\label{anotherdeform}
\end{center}
\end{table}

After the rearrangement it is not difficult to find the relation to the decomposition of the so$(8)$ representations to $[$su$(2)]^4$.
Namely, due to the decomposition of the first few so$(8)$ representations,
\begin{align}
{\bf 8_v}&\to({\bf 2},{\bf 2},{\bf 1},{\bf 1})+({\bf 1},{\bf 1},{\bf 2},{\bf 2}),
\nonumber\\
{\bf 8_s}&\to({\bf 2},{\bf 1},{\bf 2},{\bf 1})+({\bf 1},{\bf 2},{\bf 1},{\bf 2}),
\nonumber\\
{\bf 8_c}&\to({\bf 2},{\bf 1},{\bf 1},{\bf 2})+({\bf 2},{\bf 1},{\bf 1},{\bf 2}),
\nonumber\\
{\bf 28}&\to({\bf 3},{\bf 1},{\bf 1},{\bf 1})+({\bf 1},{\bf 3},{\bf 1},{\bf 1})
+({\bf 1},{\bf 1},{\bf 3},{\bf 1})+({\bf 1},{\bf 1},{\bf 1},{\bf 3})
+({\bf 2},{\bf 2},{\bf 2},{\bf 2}),
\end{align}
we can successfully identify the u$(1)$ charge as the Cartan subalgebra of the last su$(2)$.

From this identification of the u$(1)$ charge we can introduce another character with two parameters and describe the worldsheet and membrane instantons as
\begin{align}
J^{\text{WS}}(\mu_\text{eff})=\sum_{m=1}^\infty d_m e^{-m\frac{\mu_\text{eff}}{k}},\quad 
J^{\text{MB}}(\mu_\text{eff})=\sum_{\ell=1}^\infty ({\widetilde b}_\ell\mu_\text{eff}+{\widetilde c}_\ell)e^{-\ell\mu_\text{eff}},
\end{align}
where the instanton coefficients are given by
\begin{align}
d_m&=\sum_{nd=m}\frac{(-1)^m}{n}\sum_{j_\text{L},j_\text{R}}
n_{j_\text{L},j_\text{R}}^{d,{\bf R}}\frac{(-1)^{d-1}}{(2\sin\frac{\pi n}{k})^2}
\chi_{\bf R}(e^{-\pi inb_\text{I}},e^{-\pi inb_\text{II}})
\chi_{j_\text{L}}(e^{\frac{2\pi ni}{k}})
\chi_{j_\text{R}}(1),\nonumber\\
{\widetilde b}_\ell&=\sum_{nd=\ell}
\frac{1}{n}\sum_{j_\text{L},j_\text{R}}n_{j_\text{L},j_\text{R}}^{d,{\bf R}}
\frac{(-1)^{d}d}{4\pi \sin \pi nk}
\chi_{\bf R}(e^{-\pi inkb_\text{I}},e^{-\pi inkb_\text{II}})
\chi_{j_\text{L}}(e^{\pi ink})
\chi_{j_\text{R}}(e^{\pi ink}),\nonumber\\
{\widetilde c}_\ell&=-k^2\frac{\partial}{\partial k}
\biggl[\frac{{\widetilde b}_\ell}{\ell k}\biggr]_{b_\text{I},b_\text{II}},
\end{align}
with $(b_\text{I},b_\text{II})=(1-M_\text{I}/k,-2M_\text{II}/k)$.
Note that in the coefficient ${\widetilde c}_\ell$ we treat $b_\text{I}$ and $b_\text{II}$ to be independent of $k$ under the derivative.
Using the representations of $\text{so}(10)$ in table \ref{BPS}, table \ref{BPS67}, table \ref{BPS8} and the characters in appendix \ref{chiIandII}, we find that this simple expression reproduces all the instanton coefficients listed in appendix C of \cite{MNN} for $1\le d\le 8$.

\section{Orbifold $(2,1)$ model}\label{2121}

There is one more interesting theory which is solvable from the group-theoretical viewpoint.
One lesson we learned from the study of the superconformal Chern-Simons theory with the orthosymplectic gauge group in \cite{MS1} (see also \cite{HondaOSp,MS2,OkuyamaOSp,MN5}) is that sometimes the duplicate quiver is easier than the original one.
In the previous sections we have struggled for expressing the reduced grand potential of the $(2,1)$ model in terms of the free energy of topological strings \eqref{np}.
Here instead let us consider the duplicate $(2,1,2,1)$ model, which is the U$(N)^6$ superconformal Chern-Simons theory with $\{s_a\}=\{+1,+1,-1,+1,+1,-1\}$.
The physical interpretation of the repetition of $\{s_a\}$ is the orbifold \cite{HLLLP1,IK1} and we often refer to the $(2,1,2,1)$ model also as the ${\mathbb Z}_2$ orbifold of the $(2,1)$ model.
Since the odd membrane instantons of the $(2,1)$ model \eqref{21mbexpand} look very similar to those of the orthosymplectic theory \cite{MS1}, it is natural to expect that the odd membrane instantons are projected out in the duplicate $(2,1,2,1)$ model as well and the reduced grand potential falls into the standard expression \eqref{np} easily.

Before starting the computation of the instantons in the $(2,1,2,1)$ model, let us guess which set of the BPS indices should govern the model.
From the Newton polygon, the general deformation of the $(2,1,2,1)$ model corresponds to a genus-three curve, which seems not so easy from the current technology.
However, as explained carefully in \cite{BBT,KY} (see figure 8 in \cite{KY}), the $E_7$ del Pezzo geometry also appears as a special case of the same curve with the parameters tuned (which reduces the curve to genus-one).\footnote{
We can check explicitly that, in the classical limit $k\rightarrow 0$, the genus of the curve degenerates due to the singularity of the curve.
We thank Yasuhiko Yamada for valuable discussions.
}
Hence, we expect that the $(2,1,2,1)$ model is governed by the BPS indices of the local $E_7$ del Pezzo geometry.

\subsection{Instantons}

The reduced grand potential of the ${\mathbb Z}_2$ orbifold theory $J^{[2]}(\mu)$ can be obtained from the original one $J^{[1]}(\mu)$ by applying the rule of \cite{HM},
\begin{align}
J^{[2]}(\mu)&=J^{[1]}\Bigl(\frac{\mu+\pi i}{2}\Bigr)
+J^{[1]}\Bigl(\frac{\mu-\pi i}{2}\Bigr)
\nonumber\\
&\quad
+\log\biggl[1+\sum_{n\ne 0}
e^{J^{[1]}(\frac{\mu+\pi i}{2}+2\pi in)+J^{[1]}(\frac{\mu-\pi i}{2}-2\pi in)
-J^{[1]}(\frac{\mu+\pi i}{2})-J^{[1]}(\frac{\mu-\pi i}{2})}\biggr].
\label{orb}
\end{align}
Since we shall discuss both the $(2,1)$ model and the $(2,1,2,1)$ model, to avoid confusions, we put the superscripts $(2,1)$ and $(2,1,2,1)$ to each quantity in this subsection to denote which model the quantity is associated to.

When we consider the duplicate model in \eqref{orb}, we need to substitute $(\mu\pm\pi i)/2\pm 2\pi in$ for $\mu$.
The reduced grand potential $J^{(2,1)}(\mu)$ \eqref{J21} depends on $\mu$ only through $\mu^{(2,1)}_\text{eff}$ in \eqref{mueff21}.
Since the instanton effect in \eqref{mueff21} is simply $e^{-2\mu}$, we can define a common effective chemical potential $\mu^{(2,1,2,1)}_\text{eff}$ for the $(2,1,2,1)$ model
\begin{align}
\frac{\mu^{(2,1,2,1)}_\text{eff}}{2}
=\begin{cases}
\displaystyle
\frac{\mu}{2}+2e^{-\mu}{}_4F_3\Bigl(1,1,\frac{3}{2},\frac{3}{2};
2,2,2;-16e^{-\mu}\Bigr),&\text{ for odd }k,\\[6pt]
\displaystyle
\frac{\mu}{2}+6e^{-\mu}{}_4F_3\Bigl(1,1,\frac{7}{4},\frac{5}{4};
2,2,2;-64e^{-\mu}\Bigr),&\text{ for even }k,
\end{cases}
\label{mu2121}
\end{align}
independent of $n$ and substitute $(\mu^{(2,1,2,1)}_\text{eff}\pm\pi i)/2\pm 2\pi in$ for $\mu^{(2,1)}_\text{eff}$ in \eqref{J21}.

Note that there is a great simplification in the ``twisted'' sectors ($n\ne 0$) for the current case.
After the substitution, the exponent in the twisted sector becomes
\begin{align}
&J^{(2,1)}\Bigl(\frac{\mu+\pi i}{2}+2\pi in\Bigr)
+J^{(2,1)}\Bigl(\frac{\mu-\pi i}{2}-2\pi in\Bigr)
-J^{(2,1)}\Bigl(\frac{\mu+\pi i}{2}\Bigr)
-J^{(2,1)}\Bigl(\frac{\mu-\pi i}{2}\Bigr)
\nonumber\\&\quad
=-2n(2n+1)\pi^2C^{(2,1)}\mu^{(2,1,2,1)}_\text{eff}
-4\sum_{m=1}^\infty d^{(2,1)}_m
\sin\Bigl(\frac{(2n+1)m\pi}{k}\Bigr)\sin\Bigl(\frac{2nm\pi}{k}\Bigr)
e^{-\frac{m}{k}\mu^{(2,1,2,1)}_\text{eff}}.
\label{noMB}
\end{align}
Note that both the even and odd membrane instanton parts cancel among themselves, leaving only the perturbative part and the worldsheet instanton part.
Furthermore, if we use the result $C^{(2,1)}=1/(\pi^2k)$, the exponential function of the perturbative part becomes $e^{-\frac{2n(2n+1)}{k}\mu^{(2,1,2,1)}_\text{eff}}$, giving rise to the worldsheet instanton contribution.
Namely, after substituting the worldsheet instanton part \eqref{noMB} into the logarithmic function in \eqref{orb}, we find that the twisted sectors only give the worldsheet instanton for the current case.

Therefore, the membrane instanton part comes directly from the ``untwisted'' sector ($n=0$), as was the case for the perturbative part \cite{HM}.
Again, the contributions from the odd membrane instantons $e^{-(2\ell-1)\mu^{(2,1)}_\text{eff}}$ of the $(2,1)$ model disappear and the even membrane instantons $e^{-2\ell\mu^{(2,1)}_\text{eff}}$ give rise to the standard expression of the membrane instantons $e^{-\ell\mu^{(2,1,2,1)}_\text{eff}}$.
Finally, we find that the reduced grand potential of the $(2,1,2,1)$ model becomes
\begin{align}
J^{(2,1,2,1)}(\mu)
&=J^{(2,1,2,1)\text{pert}}(\mu^{(2,1,2,1)}_\text{eff})
+J^{(2,1,2,1)\text{np}}(\mu^{(2,1,2,1)}_\text{eff}),\nonumber\\
J^{(2,1,2,1)\text{np}}(\mu^{(2,1,2,1)}_\text{eff})
&=J^{(2,1,2,1)\text{WS}}(\mu^{(2,1,2,1)}_\text{eff})
+J^{(2,1,2,1)\text{MB}}(\mu^{(2,1,2,1)}_\text{eff}),
\end{align}
with each part given by
\begin{align}
J^{(2,1,2,1)\text{pert}}(\mu^{(2,1,2,1)}_\text{eff})
&=\frac{C^{(2,1,2,1)}}{3}(\mu^{(2,1,2,1)}_{\rm eff})^3
+B^{(2,1,2,1)}\mu^{(2,1,2,1)}_{\rm eff}+A^{(2,1,2,1)},
\nonumber\\
J^{(2,1,2,1)\text{WS}}(\mu^{(2,1,2,1)}_\text{eff})
&=\sum_{m=1}^\infty d^{(2,1,2,1)}_m
e^{-\frac{m}{k}\mu^{(2,1,2,1)}_{\rm eff}},
\nonumber\\
J^{(2,1,2,1)\text{MB}}(\mu^{(2,1,2,1)}_\text{eff})
&=\sum_{\ell=1}^\infty
(\widetilde{b}^{(2,1,2,1)}_{\ell}\mu^{(2,1,2,1)}_{\rm eff}
+\widetilde{c}^{(2,1,2,1)}_{\ell})
e^{-\ell\mu^{(2,1,2,1)}_{\rm eff}}.
\label{J2121}
\end{align}
Here the coefficients of the perturbative part are given by
\begin{align}
C^{(2,1,2,1)}=\frac{1}{4\pi^2k},\quad
B^{(2,1,2,1)}=-\frac{1}{3k}+\frac{k}{12},
\end{align}
while those of the membrane instantons are given by
\begin{align}
\widetilde{b}^{(2,1,2,1)}_{\ell}
=(-1)^\ell\widetilde{b}^{(2,1)}_{2\ell},\quad
\widetilde{c}^{(2,1,2,1)}_{\ell}
=2(-1)^\ell\widetilde{c}^{(2,1)}_{2\ell},
\label{bc}
\end{align}
which indicates the derivative relation
\begin{align}
\widetilde{c}^{(2,1,2,1)}_{\ell}
=-k^2\frac{d}{dk}\frac{\widetilde{b}^{(2,1,2,1)}_{\ell}}{\ell k}.
\end{align}

The coefficients of the worldsheet instantons $d_m^{(2,1,2,1)}$ are obtained by multiplying $d^{(2,1)}_m$ with the cosine factor $2\cos m\pi/k$ coming from the substitution of $(\mu^{(2,1,2,1)}_\text{eff}\pm\pi i)/2$ \eqref{mu2121} and also taking into account the twisted sector \eqref{noMB} with $n$ satisfying $2n(2n+1)\le m$.
The explicit relations for the first few coefficients are given as
\begin{align}
d_1^{(2,1,2,1)}&=2\cos\Bigl(\frac{\pi}{k}\Bigr)d_1^{(2,1)},\quad
d_2^{(2,1,2,1)}=2\cos\Bigl(\frac{2\pi}{k}\Bigr)d_2^{(2,1)}+1,\nonumber\\
d_3^{(2,1,2,1)}&=2\cos\Bigl(\frac{3\pi}{k}\Bigr)d_3^{(2,1)}
-4\sin\Bigl(\frac{\pi}{k}\Bigr)\sin\Bigl(\frac{2\pi}{k}\Bigr)d_1^{(2,1)},\nonumber\\
d_4^{(2,1,2,1)}&=2\cos\Bigl(\frac{4\pi}{k}\Bigr)d_4^{(2,1)}
-4\sin\Bigl(\frac{2\pi}{k}\Bigr)\sin\Bigl(\frac{4\pi}{k}\Bigr)d_2^{(2,1)}
+8\sin^2\Bigl(\frac{\pi}{k}\Bigr)\sin^2\Bigl(\frac{2\pi}{k}\Bigr)(d_1^{(2,1)})^2-\frac{1}{2},
\nonumber\\
&\cdots.
\label{d2121_d21example}
\end{align}

To summarize, we obtain the membrane instanton coefficients in the $(2,1,2,1)$ model of degree $\ell$ directly from those in the $(2,1)$ model of degree $2\ell$ using \eqref{bc}, while for the worldsheet instanton coefficients of degree $m$ we need to expand \eqref{orb} up to the $m$-th order with the help of \eqref{noMB} as in \eqref{d2121_d21example}.
Hereafter we shall only discuss the $(2,1,2,1)$ model and omit the superscript $(2,1,2,1)$.

\subsection{Characters}

Due to the difference in the odd instantons and the even instantons, we adopt an alternating multi-covering structure motivated by \eqref{alternating}.
\begin{align}
d_m=\sum_{n|m}\frac{1}{n}\biggl(\delta^+_{\frac{m}{n}}\Bigl(\frac{k}{n}\Bigr)
+(-1)^{n}\delta^-_{\frac{m}{n}}\Bigl(\frac{k}{n}\Bigr)\biggr).
\end{align}
By comparing with the result obtained in \eqref{orb}, we find that the first few instanton coefficients are given by
\begin{align}
&\delta^+_1(k)-\delta^-_1(k)=\frac{8}{(2\sin\frac{\pi}{k})^2},\quad
\delta^+_1(k)+\delta^-_1(k)=\frac{56}{(2\sin\frac{\pi}{k})^2},\nonumber\\
&\delta^+_2(k)-\delta^-_2(k)=-\frac{16}{(2\sin\frac{\pi}{k})^2}+3,
\end{align}
which gives
\begin{align}
&\delta^+_1(k)=\frac{32}{(2\sin\frac{\pi}{k})^2},\quad
\delta^-_1(k)=\frac{24}{(2\sin\frac{\pi}{k})^2}.
\label{delta1}
\end{align}
It is then interesting to compare these coefficients with the tables for the BPS indices of the local $E_7$ del Pezzo geometry in \cite{HKP}.
Let us decompose the $E_7$ representation ${\bf 56}$ appearing in $d=1$ to the subalgebra so$(12)\times$su$(2)$,
\begin{align}
{\bf 56}&\to({\bf 12},{\bf 2})+({\bf 32},{\bf 1}),
\end{align}
and identify $32$ and $24$ in the numerator in \eqref{delta1} respectively as $32\times 1$ and $12\times 2$.
We can imagine that the even(bosonic) and odd(fermionic) representations in the congruency class of su$(2)$ contribute to $\delta^+_d(k)$ and $\delta^-_d(k)$ respectively.

Namely, once the BPS indices in the tables of \cite{HKP} are partitioned into the $E_7$ representations
\begin{align}
N^d_{j_\text{L},j_\text{R}}=(-1)^{d-1}\sum_{\bf R}n^{d,{\bf R}}_{j_\text{L},j_\text{R}}
\dim({\bf R}),
\end{align}
we propose that the multi-covering component of the worldsheet instantons is computed by
\begin{align}
\delta^\pm_d(k)&=\frac{(-1)^{d-1}}{(2\sin\frac{\pi}{k})^2}
\sum_{j_\text{L},j_\text{R}}\sum_{\bf R}
n^{d,{\bf R}}_{j_\text{L},j_\text{R}}
n^{\pm}_{\bf R}
\chi_{j_\text{L}}(e^{\frac{2\pi i}{k}})\chi_{j_\text{R}}(1),
\end{align}
where we have defined
\begin{align}
n^{+/-}_{\bf R}
=\sum_{{\bm\rho}:\text{even}/\text{odd}}
\dim{\bm\rho}\cdot\dim{\bf r},
\end{align}
for the decomposition
\begin{align}
E_7&\to\text{so}(12)\times\text{su}(2),\quad
{\bf R}\to\sum({\bf r},{\bm\rho}).
\end{align}

For the membrane instantons, we assume the standard multi-covering structure
\begin{align}
\widetilde b_\ell=\sum_{n|\ell}\frac{1}{n}\beta_{\frac{\ell}{n}}(nk),
\end{align}
though for the comparison with the worldsheet instantons we also need the parity separation 
\begin{align}
\beta_d(k)=\beta^+_d(k)+\beta^-_d(k),\quad
\beta^\pm_d(k)=\frac{(-1)^{d}d}{4\pi\sin\pi k}
\sum_{j_\text{L},j_\text{R}}\sum_{\bf R}
n^{d,{\bf R}}_{j_\text{L},j_\text{R}}
\chi^\pm_{\bf R}(e^{-\pi ik})
\chi_{j_\text{L}}(e^{\pi ik})\chi_{j_\text{R}}(e^{\pi ik}).
\end{align}
Here we have defined the $E_7$ characters $\chi_{\bf R}^\pm(q)$ from the so$(12)$ characters $\chi_{\bf r}(q)$ as
\begin{align}
\chi_{\bf R}^\pm(q)=\sum_{{\bm\rho}:\text{even}/\text{odd}}\dim{\bm\rho}\cdot\chi_{\bf r}(q),
\quad
\chi_{\bf r}(q)=\sum_{h}q^{h}\dim({\bf r'})_{h},
\end{align}
with $h$ specifying the u$(1)$ charge in the further decomposition of the so$(12)$ representations to the subalgebra so$(10)\times$u$(1)$
\begin{align}
\text{so}(12)\to\text{so}(10)\times\text{u}(1),\quad
{\bf r}\to\sum({\bf r'})_{h}.
\end{align}

\begin{table}[ht!]
\begin{center}
\begin{tabular}{|c|c|c|c|c|}
\hline
$d$&$(j_\text{L},j_\text{R})$&BPS&representations\\
\hline\hline
$1$&$(0,0)$&$56$&${\bf 56}$\\
\hline
$2$&$(0,\frac{1}{2})$&$133$&${\bf 133}$\\
\cline{2-4}
&$(\frac{1}{2},1)$&$1$&${\bf 1}$\\
\hline
$3$&$(0,1)$&$912$&${\bf 912}$\\
\cline{2-4}
&$(0,0),(\frac{1}{2},\frac{3}{2})$&$56$&${\bf 56}$\\
\hline
$4$&$(0,\frac{3}{2})$&$8778$&${\bf 8645}+{\bf 133}$\\
\cline{2-4}
&$(0,\frac{1}{2}),(\frac{1}{2},2)$&$1673$&${\bf 1539}+{\bf 133}+{\bf 1}$\\
\cline{2-4}
&$(\frac{1}{2},1)$&$134$&${\bf 133}+{\bf 1}$\\
\cline{2-4}
&$(1,\frac{5}{2})$&$133$&${\bf 133}$\\
\cline{2-4}
&$(0,\frac{5}{2}),(1,\frac{3}{2}),(\frac{3}{2},3)$&$1$&${\bf 1}$\\
\hline
$5$&$(0,2)$&$93688$&${\bf 86184}+{\bf 6480}+{\bf 912}+2\times{\bf 56}$\\
\cline{2-4}
&$(0,1),(\frac{1}{2},\frac{5}{2})$&$36080$&${\bf 27664}+{\bf 6480}+2\times{\bf 912}+2\times{\bf 56}$\\
\cline{2-4}
&$(\frac{1}{2},\frac{3}{2})$&$8472$&${\bf 6480}+2\times{\bf 912}+3\times{\bf 56}$\\
\cline{2-4}
&$(1,3)$&$7504$&${\bf 6480}+{\bf 912}+2\times{\bf 56}$\\
\cline{2-4}
&$(0,0)$&$6592$&${\bf 6480}+2\times{\bf 56}$\\
\cline{2-4}
&$(1,2)$&$1024$&${\bf 912}+2\times{\bf 56}$\\
\cline{2-4}
&$(0,3),(\frac{1}{2},\frac{1}{2}),(\frac{3}{2},\frac{7}{2})$&$968$&${\bf 912}+{\bf 56}$\\
\cline{2-4}
&$(\frac{1}{2},\frac{7}{2}),(1,1),(\frac{3}{2},\frac{5}{2}),(2,4)$&$56$&${\bf 56}$\\
\hline
\end{tabular}
\caption{The constituent representations for the total BPS indices of the $(2,1,2,1)$ model for $1\le d\le 5$.}
\label{BPSE7}
\end{center}
\end{table}

With this identification, the remaining task is to separate the BPS indices given in \cite{HKP} as the $E_7$ representations.
Fortunately, this is given explicitly in \cite{HKP} (see table \ref{BPSE7}).
Surprisingly, we can confirm that the BPS indices with the identification of the representations given in \cite{HKP} correctly reproduce the worldsheet instantons and the membrane instantons in appendix \ref{2121inst} for $1\le d\le 4$ after decomposing the $E_7$ representations to so$(10)\times$u$(1)\times$su$(2)$ using the group-theoretical results in appendix \ref{E7}.
Comparing the congruency class ${\mathbb Z}_2$ of $E_7$, it is interesting to observe that all of the representations appearing in degree $d$ belongs to the class $d$ mod $2$.

The identification of the representations for $d=5$ given in \cite{HKP}, however, does not obey the congruency class and the decomposition of the $E_7$ representations does not give the instanton effects correctly.
Hence we assume general degeneracies $n^{d=5,{\bf R}}_{j_\text{L},j_\text{R}}$ of the representations obeying the congruency class and solve the conditions to match the worldsheet instantons and the membrane instantons listed in appendix \ref{2121inst}.
We have found a unique positive solution $\{n^{d=5,{\bf R}}_{j_\text{L},j_\text{R}}\}$ given in table \ref{BPSE7}.

As in the case of the rank-deformed $(2,2)$ model studied in the previous section, we  could introduce five K\"ahler parameters
\begin{align}
T_n=\frac{\mu_\text{eff}}{k}+n\pi i,\quad (n=0,\pm 1,\pm 2).
\end{align}
There are again, however, not enough data to completely determine the split of the BPS indices.
We have chosen alternatively to express our final result with the characters.

\section{Discussions}
\label{discuss}

We have revisited the grand potential of the $(2,1)$ model.
We first observe that the worldsheet instantons of the $(2,1)$ model coincide exactly with that of the rank deformed $(2,2)$ model through the relation \eqref{22&21}.
This gives us a hint for the novel multi-covering structure of the membrane instantons \eqref{bevenmc}.
We also observe that the BPS indices for the $(2,2)$ model are those for the local $D_5$ del Pezzo geometry with the decomposition of the so$(10)$ representations to the subalgebra so$(8)\times$u$(1)$.
With these observations in mind, we are able to construct a framework to reproduce the multi-covering structure \eqref{wsmc}, \eqref{bevenmc}, \eqref{coddmc} and the derivative relations \eqref{derivative} for the reduced grand potential of the $(2,1)$ model using the topological string free energy by introducing the four K\"ahler parameters \eqref{4K}.
After identifying the BPS indices, we discover that the BPS indices are those obtained by further decomposing the so$(10)$ representations to so$(6)\times$u$(1)\times$u$(1)$.
We also explain that it is natural that the same set of the BPS indices is used for both the $(2,2)$ model and the $(2,1)$ model from the viewpoint of the Newton polygon.

We have continued to study the $(2,1,2,1)$ model, which is the ${\mathbb Z}_2$ orbifold of the $(2,1)$ model, and find that this time the BPS indices are those of the local $E_7$ del Pezzo geometry, with the $E_7$ representations decomposed to the subalgebra so$(10)\times$u$(1)\times$su$(2)$.
Though we have not been able to identify the correct representations for the local $E_7$ del Pezzo geometry in $d=6$ so far, we have listed the worldsheet instanton and the membrane instanton in appendix \ref{2121inst} so that it can be checked in the future.

From the viewpoint of five-dimensional gauge theories \cite{Seiberg}, the local $D_5$ del Pezzo geometry and the local $E_7$ del Pezzo geometry are respectively associated to the $\mathcal N=1$ SU$(2)$ Yang-Mills theories with $N_f=4$ and $N_f=6$ matters, possessing the perturbative flavor symmetries so$(8)$ and so$(12)$.
It is only after we include the non-perturbative effects that the flavor symmetries are enhanced to $D_5$ and $E_7$.
This may explain why we first consider the decomposition of the so$(10)$ representations to so$(8)\times$u$(1)$ and that of the $E_7$ representations to so$(12)\times$su$(2)$ when studying the instantons.
Then, it remains to see which Weyl symmetries the models or the deformations preserve.
It would be interesting to figure out the general rule to identify the u$(1)$ charges.

In our determination of the representations, we have observed that the representations utilized in the BPS indices of degree $d$ are all in the congruency class $d$.
We would like to know how this can be proved mathematically rigorously.

Years ago it was difficult to find the expression of the $(2,1)$ model and its cousins.
We believe that our work has opened up a new avenue towards more general understanding of the partition function of the ${\mathcal N}=4$ superconformal Chern-Simons theories.
We would like to pursue more examples, such as the $(p,q)$ models, for a concrete view of the non-perturbative effects.
 
From the above several examples along with those in \cite{MN2,HHO}, the description of the non-perturbative effects of the reduced grand potential using the topological string theory \eqref{np} seems to work at least for the genus-one curve.
For a general $(p,q)$ model the Newton polygon suggests the curve to be of higher genus, hence it is desired to know what the correct description is for higher genus curves.
Especially we would like to see explicitly how recent proposals on the spectral determinant of higher genus curves \cite{CGM1,CGM2} works for these superconformal Chern-Simons theories.
Our orbifold $(2,1)$ model may be instructive in the sense that on one hand the associated curve is generally of genus-three, though on the other hand the curve degenerates to genus-one.

\appendix

\section{Data for $(2,1)$/$(2,2)$ model and $D_5$}\label{D5data}

In this appendix we summarize the data which are relevant in discussing the relation between the instanton effects of the $(2,1)$/$(2,2)$ models and the free energy of the topological string theory on the local $D_5$ del Pezzo geometry.
In appendix \ref{21instanton} we display the instanton coefficients of the $(2,1)$ model in terms of the multi-covering components.
In appendix \ref{irreps10} we list the irreducible representations of $\text{so}(10)$ and the characters with a single $\text{u}(1)$ fugacity associated to the decomposition of the so$(10)$ representations to the subalgebra so$(8)\times$u$(1)$.
These data are used to determine the representations which the BPS indices of the local $D_5$ del Pezzo geometry consist of from the instanton coefficients of the $(2,2)$ model.
In appendix \ref{aw} we list the characters with an additional $\text{u}(1)$ fugacity associated with the further decomposition of the so$(8)$ representations to the subalgebra so$(6)\times$u$(1)$, which appear in the instanton coefficients of the $(2,1)$ model.
Finally, in appendix \ref{chiIandII} we turn on the second fugacity in a different way so that the characters reproduce the instanton coefficients of the $(2,2)$ model with the gauge group $\text{U}(N+M_\text{II})_k\times \text{U}(N+M_\text{I})_0\times \text{U}(N+2M_\text{I}+M_\text{II})_{-k}\times \text{U}(N+M_\text{I})_0$.

\subsection{Instanton coefficients for $(2,1)$ model}\label{21instanton}

We shall list the explicit form of the instanton coefficients for the $(2,1)$ model.
The first several coefficients of the worldsheet instantons are given by
\begin{align}
\delta_1(k)&=\frac{4\cos\frac{\pi}{k}}{\sin^2\frac{2\pi}{k}},\nonumber\\
\delta_2(k)&=-\frac{4+\cos\frac{2\pi}{k}}{\sin^2\frac{2\pi}{k}},\nonumber\\
\delta_3(k)&=\frac{12\cos\frac{\pi}{k}}{\sin^2\frac{2\pi}{k}},\nonumber\\
\delta_4(k)&=-\frac{32+16\cos\frac{2\pi}{k}}{\sin^2\frac{2\pi}{k}}+5,\nonumber\\
\delta_5(k)&=\frac{220\cos\frac{\pi}{k}+20\cos\frac{3\pi}{k}}
{\sin^2\frac{2\pi}{k}}-96\cos\frac{\pi}{k},\nonumber\\
\delta_6(k)&=-\frac{780+579\cos\frac{2\pi}{k}}{\sin^2\frac{2\pi}{k}}
+\biggl(848+480\cos\frac{2\pi}{k}\biggr)
-\biggl(256+64\cos\frac{2\pi}{k}\biggr)\sin^2\frac{2\pi}{k},\nonumber\\
\delta_7(k)&=\frac{7168\cos\frac{\pi}{k}+1260\cos\frac{3\pi}{k}}
{\sin^2\frac{2\pi}{k}}
-\biggl(13232\cos\frac{\pi}{k}+1696\cos\frac{3\pi}{k}\biggr)\nonumber\\
&\qquad+\biggl(9472\cos\frac{\pi}{k}+576\cos\frac{3\pi}{k}\biggr)\sin^2\frac{2\pi}{k}
-2560\cos\frac{\pi}{k}\sin^4\frac{2\pi}{k}
\label{21ws1to7}
\end{align}
while the odd membrane instantons are
\begin{align}
\gamma_1(k)&=-\frac{\sin\pi k}{\sin^2\frac{\pi k}{2}},\nonumber\\
\gamma_3(k)&=-\frac{\sin\pi k+\sin 2\pi k}{\sin^2\frac{\pi k}{2}},\nonumber\\
\gamma_5(k)&=-\frac{2\sin\pi k+6\sin 2\pi k+6\sin 3\pi k+2\sin 4\pi k}
{\sin^2\frac{\pi k}{2}},\nonumber\\
\gamma_7(k)&=-\bigl(13\sin\pi k+38\sin 2\pi k+68\sin 3\pi k+68\sin 4\pi k
+38\sin 5\pi k+13\sin 6\pi k
\nonumber\\&\hspace{5mm}
+2\sin 7\pi k\bigr)\big/\bigl(\sin^2{\textstyle\frac{\pi k}{2}}\bigr),\nonumber\\
\gamma_9(k)&=-\bigl(150\sin\pi k+397\sin 2\pi k+754\sin 3\pi k+1053\sin 4\pi k
+1053\sin 5\pi k
\nonumber\\&\hspace{5mm}
+754\sin 6\pi k+399\sin 7\pi k+164\sin 8\pi k+52\sin 9\pi k+14\sin 10\pi k
+2\sin 11\pi k\bigr)
\nonumber\\&\hspace{5mm}
\big/\bigl(\sin^2{\textstyle\frac{\pi k}{2}}\bigr),\nonumber\\
\gamma_{11}(k)&=-\bigl(2469\sin\pi k+5880\sin 2\pi k+10694\sin 3\pi k
+16180\sin 4\pi k+20090\sin 5\pi k
\nonumber\\&\hspace{5mm}
+20092\sin 6\pi k+16194\sin 7\pi k+10751\sin 8\pi k+6064\sin 9\pi k+3002\sin 10\pi k
\nonumber\\&\hspace{5mm}
+1328\sin 11\pi k+533\sin 12\pi k+184\sin 13\pi k+57\sin 14\pi k+14\sin 15\pi k
\nonumber\\&\hspace{5mm}
+2\sin 16\pi k\bigr)\big/\bigl(\sin^2{\textstyle\frac{\pi k}{2}}\bigr),
\label{21mbodd}
\end{align}
and the even membrane instantons are
\begin{align}
\beta'_2(k)&=\frac{4\sin\pi k+\sin 2\pi k}{2\pi\sin^2\frac{\pi k}{2}},\nonumber\\
\beta'_4(k)&=\frac{5\sin\pi k+6\sin 2\pi k+5\sin 3\pi k}{\pi\sin^2\frac{\pi k}{2}},
\nonumber\\
\beta'_6(k)&=\frac{3(14\sin\pi k+28\sin 2\pi k+48\sin 3\pi k+28\sin 4\pi k
+14\sin 5\pi k+\sin 6\pi k)}{2\pi\sin^2\frac{\pi k}{2}},\nonumber\\
\beta'_8(k)&=4\bigl(43\sin\pi k+98\sin 2\pi k+192\sin 3\pi k+214\sin 4\pi k
+192\sin 5\pi k+98\sin 6\pi k
\nonumber\\&\hspace{5mm}
+46\sin 7\pi k+10\sin 8\pi k+3\sin 9\pi k\bigr)
\big/\bigl(\pi\sin^2{\textstyle\frac{\pi k}{2}}\bigr),\nonumber\\
\beta'_{10}(k)&=5\bigl(904\sin\pi k+2080\sin 2\pi k+3892\sin 3\pi k+5416\sin 4\pi k
+6328\sin 5\pi k
\nonumber\\&\hspace{5mm}
+5417\sin 6\pi k+3906\sin 7\pi k+2119\sin 8\pi k+1068\sin 9\pi k+400\sin 10\pi k
\nonumber\\&\hspace{5mm}
+164\sin 11\pi k+39\sin 12\pi k+14\sin 13\pi k+\sin 14\pi k\bigr)\big/\bigl(2\pi\sin^2{\textstyle\frac{\pi k}{2}}\bigr),
\nonumber\\
\beta'_{12}(k)&=3\bigl(13269\sin\pi k+29510\sin 2\pi k+51947\sin 3\pi k
+76500\sin 4\pi k+99103\sin 5\pi k
\nonumber\\&\hspace{5mm}
+106846\sin 6\pi k+99191\sin 7\pi k
+76740\sin 8\pi k+52699\sin 9\pi k+31238\sin 10\pi k
\nonumber\\&\hspace{5mm}
+17459\sin 11\pi k
+8580\sin 12\pi k+4190\sin 13\pi k+1728\sin 14\pi k+752\sin 15\pi k
\nonumber\\&\hspace{5mm}
+240\sin 16\pi k+94\sin 17\pi k+20\sin 18\pi k+6\sin 19\pi k\bigr)
\big/\bigl(\pi\sin^2{\textstyle\frac{\pi k}{2}}\bigr).
\label{21mbeven}
\end{align}
The auxiliary membrane instantons borrowed from the $(2,2)$ model by replacing $k$ by $k/2$ \eqref{b21fromb22} are given as
\begin{align}
\beta_1(k)&=-\frac{2\sin\pi k}{\pi\sin^2\frac{\pi k}{2}},\nonumber\\
\beta_2(k)&=\frac{8\sin\pi k+\sin 2\pi k}{2\pi\sin^2\frac{\pi k}{2}},\nonumber\\
\beta_3(k)&=-\frac{6\sin\pi k+6\sin 2\pi k}{\pi\sin^2\frac{\pi k}{2}},\nonumber\\
\beta_4(k)&=\frac{9\sin\pi k+30\sin 2\pi k+9\sin 3\pi k}{\pi\sin^2\frac{\pi k}{2}},\nonumber\\
\beta_5(k)&=-\frac{20\sin\pi k+100\sin 2\pi k+100\sin 3\pi k+20\sin 4\pi k}
{\pi\sin^2\frac{\pi k}{2}}.
\label{22mb}
\end{align}
For higher instantons of the $(2,2)$ model, the function expression was not obtained from the WKB expansion
\begin{align}
\beta_6^{(2,2)}(k)&=\frac{8146}{\pi^2k}-60732k+\frac{835836\pi^2k^3}{5}
-\frac{26743288\pi^4k^5}{105}+\frac{18972788\pi^6k^7}{75}+{\mathcal O}(k^9),
\nonumber\\
\beta_7^{(2,2)}(k)&=-\frac{2890808}{49\pi^2k}+\frac{1853576k}{3}
-\frac{110179048\pi^2k^3}{45}+\frac{741506416\pi^4k^5}{135}
-\frac{5548809784\pi^6k^7}{675}
\nonumber\\&\qquad
+{\mathcal O}(k^9),
\nonumber\\
\beta_8^{(2,2)}(k)&=\frac{7168777}{16\pi^2k}-\frac{18917506k}{3}
+\frac{1543348448\pi^2k^3}{45}-\frac{14523693056\pi^4k^5}{135}
\nonumber\\&\qquad
+\frac{1083571808768\pi^6k^7}{4725}+{\mathcal O}(k^9).
\label{beta22WKB}
\end{align}

\subsection{Decomposition of so$(10)$ representations}\label{irreps10}

\begin{table}[ht!]
\begin{center}
\begin{tabular}{|c||c|c|c|c|c|}
\hline
so(10)&$0$&$\pm 1$&$\pm 2$&$\pm 3$&$\pm 4$\\\hline\hline
${\bf 1}$&${\bf 1}$&&&&
\\\hline
${\bf 10}$&${\bf 8_v}$&&${\bf 1}$&&
\\\hline
${\bf 16}$&&${\bf 8_{s/c}}$&&&
\\\hline
${\bf 45}$&${\bf 28}+{\bf 1}$&&${\bf 8_v}$&&
\\\hline
${\bf 54}$&${\bf 35_v}+{\bf 1}$&&${\bf 8_v}$&&${\bf 1}$
\\\hline
${\bf 120}$&${\bf 56_v}+{\bf 8_v}$&&${\bf 28}$&&
\\\hline
${\bf 126}$&${\bf 56_v}$&&${\bf 35_{s/c}}$&&
\\\hline
${\bf 144}$&&${\bf 56_{s/c}}+{\bf 8_{s/c}}$&&${\bf 8_{s/c}}$&
\\\hline
\end{tabular}
\caption{The decomposition of the so$(10)$ representations to the subalgebra so$(8)\times$u$(1)$.}
\label{so10rep}
\end{center}
\end{table}

In this appendix we list the decompositions of the first several so$(10)$ irreducible representations to the subalgebra so$(8)\times$u$(1)$.
These decompositions are helpful in identifying the irreducible representations which the total BPS indices listed in \cite{HKP} consist of.
We only list the first few representations necessary for the study of $1\le d\le 5$ in table \ref{so10rep}.
For higher degrees, we present the characters.
The character for a general representation ${\bf R}$ of Lie algebra $g$ with fugacities $\xi$ can be computed by the Weyl character formula
\begin{align}
\chi_{\bf R}(\xi)
=\lim_{\epsilon\rightarrow 0}
\frac{\sum_{w\in W_g}(-1)^{l(w)}e^{(\xi+\epsilon\rho,w(d_i\omega_i+\rho))}}
{\sum_{w\in W_g}(-1)^{l(w)}e^{(\xi+\epsilon\rho,w(\rho))}}.
\label{Wcformula}
\end{align}
Here $W_g$ is the Weyl group, $l(w)$ is the length of reflection $w\in W_g$ and $d_i$ is the Dynkin label of the representation ${\bf R}$ with $\omega_i$ being the fundamental weights and $\rho=\sum_i\omega_i$ being the Weyl vector.
For the current case of the algebra so$(10)$, if we choose the fundamental weights as
\begin{align}
&\omega_1=(1,0,0,0,0),\quad
\omega_2=(1,1,0,0,0),\quad
\omega_3=(1,1,1,0,0),\nonumber\\
&\omega_4=\Bigl(\frac{1}{2},\frac{1}{2},\frac{1}{2},\frac{1}{2},-\frac{1}{2}\Bigr),
\quad
\omega_5=\Bigl(\frac{1}{2},\frac{1}{2},\frac{1}{2},\frac{1}{2},\frac{1}{2}\Bigr),
\label{so10fundweight}
\end{align}
the fugacity for the u$(1)$ charge in decomposing the so$(10)$ representations to the subalgebra so$(8)\times$u$(1)$ is $\xi=(2\log q,0,0,0,0)$.

The explicit expression of the characters are given by
\begin{align}
\chi_{\bf 1}(q)&=
1,
\nonumber\\
\chi_{\bf 45}(q)&=
29+8(q^2+q^{-2}),
\nonumber\\
\chi_{\bf 54}(q)&=
36+8(q^2+q^{-2})+q^4+q^{-4},
\nonumber\\
\chi_{\bf 210}(q)&=
98+56(q^2+q^{-2}),
\nonumber\\
\chi_{\bf 660}(q)&=
330+120(q^2+q^{-2})+36(q^4+q^{-4})+8(q^6+q^{-6})+q^8+q^{-8},
\nonumber\\
\chi_{\bf 770}(q)&=
364+168(q^2+q^{-2})+35(q^4+q^{-4}),
\nonumber\\
\chi_{\bf 945}(q)&=
441+224(q^2+q^{-2})+28(q^4+q^{-4}),
\nonumber\\
\chi_{\bf 1050}(q)&=
420+280(q^2+q^{-2})+35(q^4+q^{-4}),
\nonumber\\
\chi_{\bf 1386}(q)&=
666+288(q^2+q^{-2})+64(q^4+q^{-4})+8(q^6+q^{-6}),
\nonumber\\
\chi_{\bf 2772}(q)&=
840+672(q^2+q^{-2})+294(q^4+q^{-4}),
\nonumber\\
\chi_{\bf 4125}(q)&=
1525+1000(q^2+q^{-2})+300(q^4+q^{-4}),
\end{align}
for the congruency class $d\equiv 0$ mod 4,
\begin{align}
\chi_{\bf 10}(q)&=
8+q^2+q^{-2},
\nonumber\\
\chi_{\bf 120}(q)&=
64+28(q^2+q^{-2}),
\nonumber\\
\chi_{\bf 126}(q)&=
56+35(q^2+q^{-2}),
\nonumber\\
\chi_{\bf 210'}(q)&=
120+36(q^2+q^{-2})+8(q^4+q^{-4})+q^6+q^{-6},
\nonumber\\
\chi_{\bf 320}(q)&=
176+64(q^2+q^{-2})+8(q^4+q^{-4}),
\end{align}
for the congruency class $d\equiv 2$ mod 4 and
\begin{align}
\chi_{\bf 16}(q)&=
8(q+q^{-1}),
\nonumber\\
\chi_{\bf 144}(q)&=
64(q+q^{-1})+8(q^3+q^{-3}),
\nonumber\\
\chi_{\bf 560}(q)&=
224(q+q^{-1})+56(q^3+q^{-3}),
\nonumber\\
\chi_{\bf 672}(q)&=
224(q+q^{-1})+112(q^3+q^{-3}),
\nonumber\\
\chi_{\bf 720}(q)&=
288(q+q^{-1})+64(q^3+q^{-3})+8(q^5+q^{-5}),
\nonumber\\
\chi_{\bf 1200}(q)&=
440(q+q^{-1})+160(q^3+q^{-3}),
\nonumber\\
\chi_{\bf 1440}(q)&=
496(q+q^{-1})+224(q^3+q^{-3}),
\end{align}
for the congruency class $d\equiv 1,3$ mod 4.

\subsection{BPS indices for so$(10)$ representations}\label{aw}

In the main text we have conjectured that the BPS indices appearing in the $(2,1)$ model are those obtained by decomposing the so$(10)$ representations to the subalgebra so$(6)\times$u$(1)\times$u$(1)$.
Then only a few combinations of the original BPS indices $N_{j_\text{L},j_\text{R}}^{(d,d_\text{w},d_\text{m})}$, called the alternating BPS indices $N'^{(d,d_\text{m})}_{j_\text{L},j_\text{R}}$ \eqref{alternating} and the weighted BPS indices $M^{(d,d_\text{m})}_{j_\text{L},j_\text{R}}$ \eqref{weighted} appear in the membrane instanton of the $(2,1)$ model.
Hence, in this appendix, we shall compute these indices for various so$(10)$ representations.

\begin{table}[ht!]
\begin{center}
\begin{tabular}{|c||c|c|c|c|}
\hline
$N'^{(d,d_\text{m})}_{j_\text{L},j_\text{R}}$&$0$&$\pm 2$&$\pm 4$&$\pm 6$\\\hline\hline
${\bf 1}$&$1$&&&\\\hline
${\bf 45}$&$5$&$4$&&\\\hline
${\bf 54}$&$12$&$4$&$1$&\\\hline
${\bf 210}$&$-6$&$-4$&&\\\hline
${\bf 770}$&$36$&$20$&$11$&\\\hline
${\bf 945}$&$9$&$16$&$4$&\\\hline
${\bf 1050}$&$-20$&$-20$&$-5$&\\\hline
${\bf 1386}$&$66$&$48$&$16$&$4$\\\hline
\end{tabular}\quad
\begin{tabular}{|c||c|c|c|}
\hline
$N'^{(d,d_\text{m})}_{j_\text{L},j_\text{R}}$&$0$&$\pm 2$&$\pm 4$\\\hline\hline
${\bf 10}$&$4$&$1$&\\\hline
${\bf 120}$&$0$&$4$&\\\hline
${\bf 126}$&$-4$&$-5$&\\\hline
${\bf 320}$&$24$&$16$&$4$\\\hline
\end{tabular}\qquad
\begin{tabular}{|c||c|c|c|}
\hline
$M^{(d,d_\text{m})}_{j_\text{L},j_\text{R}}$&$\pm 1$&$\pm 3$&$\pm 5$\\\hline\hline
${\bf 16}$&$8$&&\\\hline
${\bf 144}$&$32$&$8$&\\\hline
${\bf 560}$&$32$&$24$&\\\hline
${\bf 720}$&$96$&$32$&$8$\\\hline
\end{tabular}
\caption{(Left two)
The alternating BPS indices $(-1)^{d-1}N'^{(d,d_\text{m})}_{j_\text{L},j_\text{R}}$ for the representations of so$(10)$ in the congruency class $d\equiv 0$ or $d\equiv 2$ mod $4$ which are used in the membrane instanton effects $\beta'_d(k)$ for $d=2,4,6,8$.
(Right one)
The weighted BPS indices $(-1)^{d-1}M^{(d,d_\text{m})}_{j_\text{L},j_\text{R}}$ for the representations of so$(10)$ in the congruency class $d\equiv 1$ or $d\equiv 3$ mod $4$ which are used in the membrane instanton effects $\gamma_d(k)$ for $d=1,3,5,7$.
}
\label{Ntable}
\end{center}
\end{table}

These BPS indices can be computed from the characters with two fugacities indicating the two u$(1)$ charges in decomposing the so$(10)$ representations to the subalgebra so$(6)\times$u$(1)\times$u$(1)$.
With the same choice of the fundamental weights \eqref{so10fundweight}, the characters can be obtained by substituting $\xi=(2\log q,2\log p,0,0,0)$ into \eqref{Wcformula}, where we omit the characters of some so$(10)$ representations which are not used for the BPS indices.
Then, using \eqref{BPSch}, the two combinations of the BPS indices are obtained in table \ref{Ntable} from the characters.

The characters are given by
\begin{align}
\chi_{\bf 1}(p,q)&=
1,
\nonumber\\
\chi_{\bf 45}(p,q)&=
17+6(q^2+q^{-2})+(p^2+p^{-2})(6+q^2+q^{-2}),
\nonumber\\
\chi_{\bf 54}(p,q)&=
22+6(q^2+q^{-2})+q^4+q^{-4}+(p^2+p^{-2})(6+q^2+q^{-2})+p^4+p^{-4},
\nonumber\\
\chi_{\bf 210}(p,q)&=
46+26(q^2+q^{-2})+(p^2+p^{-2})(26+15(q^2+q^{-2})),
\nonumber\\
\chi_{\bf 770}(p,q)&=
158+82(q^2+q^{-2})+21(q^4+q^{-4})
+(p^2+p^{-2})(82+37(q^2+q^{-2})+6(q^4+q^{-4}))
\nonumber\\[-4pt]&\hspace{-5mm}
+(p^4+p^{-4})(21+6(q^2+q^{-2})+q^4+q^{-4}),
\nonumber\\
\chi_{\bf 945}(p,q)&=
193+108(q^2+q^{-2})+16(q^4+q^{-4})
\nonumber\\[-4pt]&\hspace{-5mm}
+(p^2+p^{-2})(108+52(q^2+q^{-2})+6(q^4+q^{-4}))
+(p^4+p^{-4})(16+6(q^2+q^{-2})),
\nonumber\\
\chi_{\bf 1050}(p,q)&=
170+110(q^2+q^{-2})+15(q^4+q^{-4})
\nonumber\\[-4pt]&\hspace{-5mm}
+(p^2+p^{-2})(110+75(q^2+q^{-2})+10(q^4+q^{-4}))
+(p^4+p^{-4})(15+10(q^2+q^{-2})),
\nonumber\\
\chi_{\bf 1386}(p,q)&=
290+144(q^2+q^{-2})+38(q^4+q^{-4})+6(q^6+q^{-6})
\nonumber\\[-4pt]&\hspace{-5mm}
+(p^2+p^{-2})(144+59(q^2+q^{-2})+12(q^4+q^{-4})+q^6+q^{-6})
\nonumber\\[-4pt]&\hspace{-5mm}
+(p^4+p^{-4})(38+12(q^2+q^{-2})+q^4+q^{-4})
+(p^6+p^{-6})(6+q^2+q^{-2}),
\label{chiso10cong0}
\end{align}
for the congruency class $d\equiv 0$ mod 4,
\begin{align}
\chi_{\bf 10}(p,q)&=
6+q^2+q^{-2}+p^2+p^{-2},\nonumber\\
\chi_{\bf 120}(p,q)&=
32+16(q^2+q^{-2})+(p^2+p^{-2})(16+6(q^2+q^{-2})),\nonumber\\
\chi_{\bf 126}(p,q)&=
26+15(q^2+q^{-2})+(p^2+p^{-2})(15+10(q^2+q^{-2})),\nonumber\\
\chi_{\bf 320}(p,q)&=
88+38(q^2+q^{-2})+6(q^4+q^{-4})+(p^2+p^{-2})(38+12(q^2+q^{-2})+q^4+q^{-4})
\nonumber\\[-4pt]&\hspace{-5mm}
+(p^4+p^{-4})(6+q^2+q^{-2}),
\end{align}
for the congruency class $d\equiv 2$ mod 4 and 
\begin{align}
\chi_{\bf 16}(p,q)&=
4(p+p^{-1})(q+q^{-1}),\nonumber\\
\chi_{\bf 144}(p,q)&=
4(p+p^{-1})(q+q^{-1})[5+q^2+q^{-2}+p^2+p^{-2}],\nonumber\\
\chi_{\bf 560}(p,q)&=
4(p+p^{-1})(q+q^{-1})[11+5(q^2+q^{-2})+(p^2+p^{-2})(5+q^2+q^{-2})],\nonumber\\
\chi_{\bf 720}(p,q)&=
4(p+p^{-1})(q+q^{-1})[17+5(q^2+q^{-2})+q^4+q^{-4}
\nonumber\\[-4pt]&\hspace{-5mm}
+(p^2+p^{-2})(5+q^2+q^{-2})+p^4+p^{-4}],
\end{align}
for the congruency class $d\equiv 1,3$ mod 4.

\subsection{Characters for rank-deformed $(2,2)$ model}\label{chiIandII}

In this appendix we shall list the so$(10)$ characters with two parameters for the study of the $(2,2)$ model with the rank deformation U$(N+M_\text{II})_k\times$U$(N+M_\text{I})_0\times$U$(N+2M_\text{I}+M_\text{II})_{-k}\times$U$(N+M_\text{I})_0$.
The characters are obtained by setting $\xi=(2\log q_\text{I},\log q_\text{II},\log q_\text{II},0,0)$ in the Weyl character formula \eqref{Wcformula} for the same choice of the fundamental weights \eqref{so10fundweight}.
The characters are given explicitly by
\begin{align}
\chi_{\bf 1}(q_\text{I},q_\text{II})
&=1,
\nonumber\\
\chi_{\bf 45}(q_\text{I},q_\text{II})
&=11+8(q_\text{II}+q_\text{II}^{-1})+q_\text{II}^2+q_\text{II}^{-2}
+(q_\text{I}^2+q_\text{I}^{-2})(4+2(q_\text{II}+q_\text{II}^{-1})),
\nonumber\\
\chi_{\bf 54}(q_\text{I},q_\text{II})
&=14+8(q_\text{II}+q_\text{II}^{-1})+3(q_\text{II}^2+q_\text{II}^{-2})
+(q_\text{I}^2+q_\text{I}^{-2})(4+2(q_\text{II}+q_\text{II}^{-1}))
+q_\text{I}^4+q_\text{I}^{-4},
\nonumber\\
\chi_{\bf 210}(q_\text{I},q_\text{II})
&=36+24(q_\text{II}+q_\text{II}^{-1})+7(q_\text{II}^2+q_\text{II}^{-2})
\nonumber\\[-4pt]&\hspace{-5mm}
+(q_\text{I}^2+q_\text{I}^{-2})
(20+14(q_\text{II}+q_\text{II}^{-1})+4(q_\text{II}^2+q_\text{II}^{-2})),
\nonumber\\
\chi_{\bf 770}(q_\text{I},q_\text{II})
&=104+80(q_\text{II}+q_\text{II}^{-1})+41(q_\text{II}^2+q_\text{II}^{-2})
+8(q_\text{II}^3+q_\text{II}^{-3})+q_\text{II}^4+q_\text{II}^{-4}
\nonumber\\[-4pt]&\hspace{-5mm}
+(q_\text{I}^2+q_\text{I}^{-2})
(52+40(q_\text{II}+q_\text{II}^{-1})+16(q_\text{II}^2+q_\text{II}^{-2})
+2(q_\text{II}^3+q_\text{II}^{-3}))
\nonumber\\[-4pt]&\hspace{-5mm}
+(q_\text{I}^4+q_\text{I}^{-4})
(13+8(q_\text{II}+q_\text{II}^{-1})+3(q_\text{II}^2+q_\text{II}^{-2})),
\nonumber\\
\chi_{\bf 945}(q_\text{I},q_\text{II})
&=133+104(q_\text{II}+q_\text{II}^{-1})+42(q_\text{II}^2+q_\text{II}^{-2})
+8(q_\text{II}^3+q_\text{II}^{-3})
\nonumber\\[-4pt]&\hspace{-5mm}
+(q_\text{I}^2+q_\text{I}^{-2})
(72+54(q_\text{II}+q_\text{II}^{-1})+20(q_\text{II}^2+q_\text{II}^{-2})
+2(q_\text{II}^3+q_\text{II}^{-3}))
\nonumber\\[-4pt]&\hspace{-5mm}
+(q_\text{I}^4+q_\text{I}^{-4})
(10+8(q_\text{II}+q_\text{II}^{-1})+q_\text{II}^2+q_\text{II}^{-2}),
\nonumber\\
\chi_{\bf 1050}(q_\text{I},q_\text{II})
&=126+96(q_\text{II}+q_\text{II}^{-1})+43(q_\text{II}^2+q_\text{II}^{-2})
+8(q_\text{II}^3+q_\text{II}^{-3})
\nonumber\\[-4pt]&\hspace{-5mm}
+(q_\text{I}^2+q_\text{I}^{-2})
(84+64(q_\text{II}+q_\text{II}^{-1})+28(q_\text{II}^2+q_\text{II}^{-2})
+6(q_\text{II}^3+q_\text{II}^{-3}))
\nonumber\\[-4pt]&\hspace{-5mm}
+(q_\text{I}^4+q_\text{I}^{-4})
(13+8(q_\text{II}+q_\text{II}^{-1})+3(q_\text{II}^2+q_\text{II}^{-2})),
\nonumber\\
\chi_{\bf 1386}(q_\text{I},q_\text{II})
&=178+144(q_\text{II}+q_\text{II}^{-1})+73(q_\text{II}^2+q_\text{II}^{-2})
+24(q_\text{II}^3+q_\text{II}^{-3})+3(q_\text{II}^4+q_\text{II}^{-4})
\nonumber\\[-4pt]&\hspace{-5mm}
+(q_\text{I}^2+q_\text{I}^{-2})
(88+66(q_\text{II}+q_\text{II}^{-1})+28(q_\text{II}^2+q_\text{II}^{-2})
+6(q_\text{II}^3+q_\text{II}^{-3}))
\nonumber\\[-4pt]&\hspace{-5mm}
+(q_\text{I}^4+q_\text{I}^{-4})
(24+16(q_\text{II}+q_\text{II}^{-1})+4(q_\text{II}^2+q_\text{II}^{-2}))
+(q_\text{I}^6+q_\text{I}^{-6})
(4+2(q_\text{II}+q_\text{II}^{-1})),
\label{chiso10M1M21}
\end{align}
for the congruency class $d\equiv 0$ mod $4$,
\begin{align}
\chi_{\bf 10}(q_\text{I},q_\text{II})
&=4+2(q_\text{II}+q_\text{II}^{-1})+q_\text{I}^2+q_\text{I}^{-2},
\nonumber\\
\chi_{\bf 120}(q_\text{I},q_\text{II})
&=24+16(q_\text{II}+q_\text{II}^{-1})+4(q_\text{II}^2+q_\text{II}^{-2})
+(q_\text{I}^2+q_\text{I}^{-2})
(10+8(q_\text{II}+q_\text{II}^{-1})+q_\text{II}^2+q_\text{II}^{-2}),
\nonumber\\
\chi_{\bf 126}(q_\text{I},q_\text{II})
&=20+14(q_\text{II}+q_\text{II}^{-1})+4(q_\text{II}^2+q_\text{II}^{-2})
+(q_\text{I}^2+q_\text{I}^{-2})
(13+8(q_\text{II}+q_\text{II}^{-1})+3(q_\text{II}^2+q_\text{II}^{-2})),
\nonumber\\
\chi_{\bf 320}(q_\text{I},q_\text{II})
&=56+42(q_\text{II}+q_\text{II}^{-1})+16(q_\text{II}^2+q_\text{II}^{-2})
+2(q_\text{II}^3+q_\text{II}^{-3})
\nonumber\\[-4pt]&\hspace{-5mm}
+(q_\text{I}^2+q_\text{I}^{-2})
(24+16(q_\text{II}+q_\text{II}^{-1})+4(q_\text{II}^2+q_\text{II}^{-2}))
+(q_\text{I}^4+q_\text{I}^{-4})
(4+2(q_\text{II}+q_\text{II}^{-1})),
\label{chiso10M1M22}
\end{align}
for the congruency class $d\equiv 2$ mod $4$ and
\begin{align}
\chi_{\bf 16}(q_\text{I},q_\text{II})
&=2(q_\text{I}+q_\text{I}^{-1})(q_\text{II}^{\frac{1}{2}}
+q_\text{II}^{-\frac{1}{2}}),
\nonumber\\
\chi_{\bf 144}(q_\text{I},q_\text{II})
&=2(q_\text{I}+q_\text{I}^{-1})(q_\text{II}^{\frac{1}{2}}+q_\text{II}^{-\frac{1}{2}})
\bigl[3+q_\text{II}+q_\text{II}^{-1}+q_\text{I}^2+q_\text{I}^{-2}\bigr],
\nonumber\\
\chi_{\bf 560}(q_\text{I},q_\text{II})
&=2(q_\text{I}+q_\text{I}^{-1})(q_\text{II}^{\frac{1}{2}}+q_\text{II}^{-\frac{1}{2}})
\bigl[7+6(q_\text{II}+q_\text{II}^{-1})+q_\text{II}^2+q_\text{II}^{-2}
\nonumber\\[-4pt]&\hspace{-5mm}
+(q_\text{I}^2+q_\text{I}^{-2})(3+2(q_\text{II}+q_\text{II}^{-1}))\bigr],
\nonumber\\
\chi_{\bf 720}(q_\text{I},q_\text{II})
&=2(q_\text{I}+q_\text{I}^{-1})(q_\text{II}^{\frac{1}{2}}+q_\text{II}^{-\frac{1}{2}})
\bigl[11+6(q_\text{II}+q_\text{II}^{-1})+3(q_\text{II}^2+q_\text{II}^{-2})
\nonumber\\[-4pt]&\hspace{-5mm}
+(q_\text{I}^2+q_\text{I}^{-2})(3+2(q_\text{II}+q_\text{II}^{-1}))
+q_\text{I}^4+q_\text{I}^{-4}\bigr],
\label{chiso10M1M23}
\end{align}
for the congruency class $d\equiv 1,3$ mod $4$.

\section{Data for $(2,1,2,1)$ model and $E_7$}\label{E7data}

In this appendix we summarize the data to relate the instanton coefficients of the $(2,1,2,1)$ model with the free energy of the topological string theory on the local $E_7$ del Pezzo geometry.
In appendix \ref{2121inst} we collect the instanton coefficients, while appendix \ref{E7} provides the decompositions of the irreducible representations of $E_7$ to the subalgebra so$(12)\times$su$(2)$ and the characters of the $\text{so}(12)$ representations.

\subsection{Instanton coefficients for $(2,1,2,1)$ model}\label{2121inst}

In this appendix we list the first several instanton coefficients.
For the worldsheet coefficients, following the main text, we express them by separating into the $\pm$ parts.
Note that, purely from the numerical results of the instanton effects up to degree $d$, we only obtain the difference $\delta^+_d(k)-\delta^-_d(k)$.
The separation is obtained only after studying the instanton effects up to degree $2d$ or taking care of the tables for the BPS indices of the local del Pezzo $E_7$ geometry in \cite{HKP}.
For $\delta^+_d(k)$ we obtain
\begin{align}
\delta^+_1(k)&=\frac{32}{(2\sin\frac{\pi}{k})^2},
\nonumber\\
\delta^+_2(k)&=-\frac{144}{(2\sin\frac{\pi}{k})^2}+3,
\nonumber\\
\delta^+_3(k)&=\frac{1632}{(2\sin\frac{\pi}{k})^2}-128,
\nonumber\\
\delta^+_4(k)&=-\frac{29248}{(2\sin\frac{\pi}{k})^2}+6157
-460\biggl(2\sin\frac{\pi}{k}\biggr)^2+7\biggl(2\sin\frac{\pi}{k}\biggr)^4,
\nonumber\\
\delta^+_5(k)&=\frac{652160}{(2\sin\frac{\pi}{k})^2}-288576
+59328\biggl(2\sin\frac{\pi}{k}\biggr)^2-6336\biggl(2\sin\frac{\pi}{k}\biggr)^4
+288\biggl(2\sin\frac{\pi}{k}\biggr)^6,
\nonumber\\
\delta^+_6(k)&=-\frac{16629168}{(2\sin\frac{\pi}{k})^2}+13073657
-5292592\biggl(2\sin\frac{\pi}{k}\biggr)^2+1338304\biggl(2\sin\frac{\pi}{k}\biggr)^4
\nonumber\\&\hspace{-5mm}
-215992\biggl(2\sin\frac{\pi}{k}\biggr)^6+20969\biggl(2\sin\frac{\pi}{k}\biggr)^8
-1020\biggl(2\sin\frac{\pi}{k}\biggr)^{10}+13\biggl(2\sin\frac{\pi}{k}\biggr)^{12},
\end{align}
while for $\delta^-_d(k)$ we obtain
\begin{align}
\delta^-_1(k)&=\frac{24}{(2\sin\frac{\pi}{k})^2},
\nonumber\\
\delta^-_2(k)&=-\frac{128}{(2\sin\frac{\pi}{k})^2},
\nonumber\\
\delta^-_3(k)&=\frac{1608}{(2\sin\frac{\pi}{k})^2}-96,
\nonumber\\
\delta^-_4(k)&=-\frac{29184}{(2\sin\frac{\pi}{k})^2}+5888
-384\biggl(2\sin\frac{\pi}{k}\biggr)^2,
\nonumber\\
\delta^-_5(k)&=\frac{651680}{(2\sin\frac{\pi}{k})^2}-286320
+57552\biggl(2\sin\frac{\pi}{k}\biggr)^2
-5776\biggl(2\sin\frac{\pi}{k}\biggr)^4+216\biggl(2\sin\frac{\pi}{k}\biggr)^6,
\nonumber\\
\delta^-_6(k)&=-\frac{16626048}{(2\sin\frac{\pi}{k})^2}+13053696
-5262208\biggl(2\sin\frac{\pi}{k}\biggr)^2+1316608\biggl(2\sin\frac{\pi}{k}\biggr)^4
\nonumber\\&\hspace{-5mm}
-207232\biggl(2\sin\frac{\pi}{k}\biggr)^6+18944\biggl(2\sin\frac{\pi}{k}\biggr)^8
-768\biggl(2\sin\frac{\pi}{k}\biggr)^{10}.
\end{align}

For the membrane instanton, the derivation is more direct.
We only need to apply \eqref{bc}, rewrite into the multi-covering expression and separate into $\beta^\pm_d(k)$ according to the arguments of the sine functions in the numerators.
For $\beta^+_d(k)$ we obtain
\begin{align}
\beta^+_1(k)&=-\frac{4\sin 2\pi k}{\pi\sin^2\pi k},
\nonumber\\
\beta^+_2(k)&=\frac{50\sin 2\pi k+11\sin 4\pi k}{2\pi\sin^2\pi k},
\nonumber\\
\beta^+_3(k)&=-\frac{12(15\sin 2\pi k+15\sin 4\pi k+2\sin 6\pi k)}
{\pi\sin^2\pi k},
\nonumber\\
\beta^+_4(k)&=\frac{4(863\sin 2\pi k+1630\sin 4\pi k+869\sin 6\pi k+138\sin 8\pi k
+6\sin 10\pi k)}{2\pi\sin^2\pi k},
\nonumber\\
\beta^+_5(k)&=-40\bigl(560\sin 2\pi k+1317\sin 4\pi k+1318\sin 6\pi k
+576\sin 8\pi k
\nonumber\\&\hspace{-5mm}
+127\sin 10\pi k+16\sin 12\pi k+\sin 14\pi k\bigr)\big/
\bigl(\pi\sin^2\pi k\bigr),
\nonumber\\
\beta^+_6(k)&=3\bigl(248502\sin 2\pi k+608220\sin 4\pi k+824190\sin 6\pi k
+610860\sin 8\pi k
\nonumber\\&\hspace{-5mm}
+265298\sin 10\pi k+77619\sin 12\pi k+16796\sin 14\pi k
+2652\sin 16\pi k+280\sin 18\pi k
\nonumber\\&\hspace{-5mm}
+12\sin 20\pi k\bigr)
\big/\bigl(2\pi\sin^2\pi k\bigr),
\end{align}
while for $\beta^-_d(k)$ we obtain
\begin{align}
\beta^-_1(k)&=-\frac{9\sin\pi k+\sin 3\pi k}{2\pi\sin^2\pi k},
\nonumber\\
\beta^-_2(k)&=\frac{16\sin\pi k+16\sin 3\pi k}{\pi\sin^2\pi k},
\nonumber\\
\beta^-_3(k)&=-\frac{3(56\sin\pi k+152\sin 3\pi k+57\sin 5\pi k+\sin 7\pi k)}
{2\pi\sin^2\pi k},
\nonumber\\
\beta^-_4(k)&=\frac{4(368\sin\pi k+1392\sin 3\pi k+1392\sin 5\pi k+400\sin 7\pi k
+32\sin 9\pi k)}{2\pi\sin^2\pi k},
\nonumber\\
\beta^-_5(k)&=-5\bigl(3888\sin\pi k+15280\sin 3\pi k+23489\sin 5\pi k
+15348\sin 7\pi k+4655\sin 9\pi k
\nonumber\\&\hspace{-5mm}
+767\sin 11\pi k+68\sin 13\pi k+\sin 15\pi k\bigr)
\big/\bigl(2\pi\sin^2\pi k\bigr),
\nonumber\\
\beta^-_6(k)&=48\bigl(3503\sin\pi k+13119\sin 3\pi k+23847\sin 5\pi k
+23873\sin 7\pi k+13336\sin 9\pi k
\nonumber\\&\hspace{-5mm}
+4671\sin 11\pi k+1168\sin 13\pi k
+217\sin 15\pi k+28\sin 17\pi k+2\sin 19\pi k\bigr)\big/\bigl(\pi\sin^2\pi k\bigr).
\end{align}

\subsection{Decomposition of $E_7$ representations}\label{E7}

To identify the representations which the BPS indices consist of for the $(2,1,2,1)$ model, we need to decompose the $E_7$ representations to the subalgebra so$(12)\times$su$(2)$ and further decompose the so$(12)$ representations to the subalgebra so$(10)\times$u$(1)$.
The first several decompositions are given in table A.88 of \cite{LieART}.
Though this is not enough we can continue by the Mathematica package provided there.
For our purpose, we separate the decompositions by the congruency class.
For the even congruency class we find the decompositions
\begin{align}
{\bf 1}
&\to({\bf 1},{\bf 1}),
\nonumber\\
{\bf 133}
&\to({\bf 1},{\bf 3})+(\overline{\bf 32},{\bf 2})+({\bf 66},{\bf 1}),
\nonumber\\
{\bf 1463}
&\to({\bf 66},{\bf 1})+({\bf 77},{\bf 3})+(\overline{\bf 352},{\bf 2})+({\bf 462},{\bf 1}),
\nonumber\\
{\bf 1539}
&\to({\bf 1},{\bf 1})+(\overline{\bf 32},{\bf 2})+({\bf 66},{\bf 3})+({\bf 77},{\bf 1})+(\overline{\bf 352},{\bf 2})+({\bf 495},{\bf 1}),
\nonumber\\
{\bf 7371}
&\to({\bf 1},{\bf 1})+({\bf 1},{\bf 5})+(\overline{\bf 32},{\bf 2})+(\overline{\bf 32},{\bf 4})+({\bf 66},{\bf 3})+(\overline{\bf 462},{\bf 3})+({\bf 495},{\bf 1})+({\bf 1638},{\bf 1})
\nonumber\\[-4pt]&\hspace{-2mm}
+(\overline{\bf 1728},{\bf 2}),
\nonumber\\
{\bf 8645}
&\to
({\bf 1},{\bf 3})+(\overline{\bf 32},{\bf 2})+(\overline{\bf 32},{\bf 4})+({\bf 66},{\bf 1})+({\bf 66},{\bf 3})+(\overline{\bf 352},{\bf 2})+(\overline{\bf 462},{\bf 1})+({\bf 495},{\bf 3})
\nonumber\\[-4pt]&\hspace{-2mm}
+(\overline{\bf 1728},{\bf 2})+({\bf 2079},{\bf 1}),
\nonumber\\
{\bf 40755}
&\to
(\overline{\bf 32},{\bf 2})+({\bf 66},{\bf 1})+({\bf 66},{\bf 3})+({\bf 77},{\bf 1})+({\bf 77},{\bf 3})+2(\overline{\bf 352},{\bf 2})+(\overline{\bf 352},{\bf 4})+({\bf 462},{\bf 3}),
\nonumber\\[-4pt]&\hspace{-2mm}
+({\bf 495},{\bf 1})+({\bf 495},{\bf 3})+(\overline{\bf 1728},{\bf 2})+({\bf 2079},{\bf 1})+({\bf 2079},{\bf 3})+(\overline{\bf 2112},{\bf 2})+(\overline{\bf 4928}',{\bf 2})
\nonumber\\[-4pt]&\hspace{-2mm}
+({\bf 8085},{\bf 1}),
\end{align}
while for the odd congruency class we find the decompositions
\begin{align}
{\bf 56}
&\to
({\bf 12},{\bf 2})+({\bf 32},{\bf 1}),
\nonumber\\
{\bf 912}
&\to
({\bf 12},{\bf 2})+({\bf 32},{\bf 3})+({\bf 220},{\bf 2})+({\bf 352},{\bf 1}),
\nonumber\\
{\bf 6480}
&\to
({\bf 12},{\bf 2})+({\bf 12},{\bf 4})+({\bf 32},{\bf 1})+({\bf 32},{\bf 3})+({\bf 220},{\bf 2})+({\bf 352},{\bf 1})+({\bf 352},{\bf 3})+({\bf 560},{\bf 2})
\nonumber\\[-4pt]&\hspace{-2mm}
+({\bf 792},{\bf 2})+({\bf 1728},{\bf 1}),
\nonumber\\
{\bf 24320}
&\to
({\bf 352}',{\bf 4})+({\bf 560},{\bf 2})+({\bf 1728},{\bf 1})+({\bf 2112},{\bf 3})+({\bf 4224},{\bf 1})+({\bf 4752},{\bf 2}),
\nonumber\\
{\bf 27664}
&\to
({\bf 12},{\bf 2})+({\bf 32},{\bf 1})+({\bf 32},{\bf 3})+({\bf 220},{\bf 2})+({\bf 220},{\bf 4})+({\bf 352},{\bf 1})+({\bf 352},{\bf 3})+({\bf 560},{\bf 2})
\nonumber\\[-4pt]&\hspace{-2mm}
+({\bf 792},{\bf 2})+({\bf 1728},{\bf 3})+({\bf 2112},{\bf 1})+({\bf 4928}',{\bf 1})+({\bf 4928},{\bf 2}),
\nonumber\\
{\bf 51072}
&\to
({\bf 12},{\bf 2})+({\bf 32},{\bf 1})+({\bf 220},{\bf 2})+({\bf 352},{\bf 1})+({\bf 352}',{\bf 2})+({\bf 352},{\bf 3})+({\bf 560},{\bf 2})+({\bf 560},{\bf 4})
\nonumber\\[-4pt]&\hspace{-2mm}
+({\bf 792},{\bf 2})+({\bf 1728},{\bf 1})+({\bf 1728},{\bf 3})+({\bf 2112},{\bf 1})+({\bf 2112},{\bf 3})+({\bf 4752},{\bf 2})+({\bf 4928},{\bf 2})
\nonumber\\[-4pt]&\hspace{-2mm}
+({\bf 8800},{\bf 1}),
\nonumber\\
{\bf 86184}
&\to
({\bf 12},{\bf 2})+({\bf 12},{\bf 4})+({\bf 32},{\bf 1})+({\bf 32},{\bf 3})+({\bf 32},{\bf 5})+2({\bf 220},{\bf 2})+({\bf 220},{\bf 4})+({\bf 352},{\bf 1})
\nonumber\\[-4pt]&\hspace{-2mm}
+2({\bf 352},{\bf 3})+({\bf 560},{\bf 2})+({\bf 792},{\bf 2})+({\bf 792},{\bf 4})+({\bf 1728},{\bf 1})+({\bf 1728},{\bf 3})+(\overline{\bf 4752},{\bf 2})
\nonumber\\[-4pt]&\hspace{-2mm}
+({\bf 4928}',{\bf 1})+({\bf 4928},{\bf 2})+({\bf 4928}',{\bf 3})+({\bf 8008},{\bf 2})+({\bf 13728},{\bf 1}).
\end{align}

For the study of the membrane instantons in the $(2,1,2,1)$ model, we need to further decompose the so$(12)$ representations to the subalgebra so$(10)\times$u$(1)$.
For this purpose, the characters are helpful.
These characters can be obtained by choosing $\xi=(2\log q,0,0,0,0,0)$ in the Weyl character formula \eqref{Wcformula} if we fix the fundamental weights as
\begin{align}
\omega_1&=(1,0,0,0,0,0),\quad
\omega_2=(1,1,0,0,0,0),\quad
\omega_3=(1,1,1,0,0,0),\quad
\omega_4=(1,1,1,1,0,0),\nonumber \\
\omega_5&=\Bigl(\frac{1}{2},\frac{1}{2},\frac{1}{2},\frac{1}{2},\frac{1}{2},-\frac{1}{2}\Bigr),\quad
\omega_6=\Bigl(\frac{1}{2},\frac{1}{2},\frac{1}{2},\frac{1}{2},\frac{1}{2},\frac{1}{2}\Bigr).
\end{align}
The explicit form of the characters is given by
\begin{align}
\chi_{\bf 1}(q)&=
1,
\nonumber\\
\chi_{\bf 12}(q)&=
10+(q^2+q^{-2}),
\nonumber\\
\chi_{\bf 32}(q)&=
16(q+q^{-1}),
\nonumber\\
\chi_{\bf 66}(q)&=
46+10(q^2+q^{-2}),
\nonumber\\
\chi_{\bf 77}(q)&=
55+10(q^2+q^{-2})+(q^4+q^{-4}),
\nonumber\\
\chi_{\bf 220}(q)&=
130+45(q^2+q^{-2}),
\nonumber\\
\chi_{\bf 352}(q)&=
160(q+q^{-1})+16(q^3+q^{-3}),
\nonumber\\
\chi_{{\bf 352}'}(q)&=
220+55(q^2+q^{-2})+10(q^4+q^{-4})+(q^6+q^{-6}),
\nonumber\\
\chi_{\bf 462}(q)&=
210+126(q^2+q^{-2}),
\nonumber\\
\chi_{\bf 495}(q)&=
255+120(q^2+q^{-2}),
\nonumber\\
\chi_{\bf 560}(q)&=
340+100(q^2+q^{-2})+10(q^4+q^{-4}),
\nonumber\\
\chi_{\bf 792}(q)&=
372+210(q^2+q^{-2}),
\nonumber\\
\chi_{\bf 1287}(q)&=
715+220(q^2+q^{-2})+55(q^4+q^{-4})+10(q^6+q^{-6})+(q^8+q^{-8}),
\nonumber\\
\chi_{\bf 1638}(q)&=
870+330(q^2+q^{-2})+54(q^4+q^{-4}),
\nonumber\\
\chi_{\bf 1728}(q)&=
720(q+q^{-1})+144(q^3+q^{-3}),
\nonumber\\
\chi_{\bf 2079}(q)&=
1089+450(q^2+q^{-2})+45(q^4+q^{-4}),
\nonumber\\
\chi_{\bf 2112}(q)&=
880(q+q^{-1})+160(q^3+q^{-3})+16(q^5+q^{-5}),
\nonumber\\
\chi_{\bf 2860}(q)&=
1540+550(q^2+q^{-2})+100(q^4+q^{-4})+10(q^6+q^{-6}),
\nonumber\\
\chi_{\bf 4004}(q)&=
2002+715(q^2+q^{-2})+220(q^4+q^{-4})+55(q^6+q^{-6})+10(q^8+q^{-8})
\nonumber\\[-4pt]&\quad
+(q^{10}+q^{-10}),
\nonumber\\
\chi_{\bf 4224}(q)&=
1440(q+q^{-1})+672(q^3+q^{-3}),
\nonumber\\
\chi_{\bf 4752}(q)&=
1980+1260(q^2+q^{-2})+126(q^4+q^{-4}),
\nonumber\\
\chi_{\bf 4928}(q)&=
2288+1200(q^2+q^{-2})+120(q^4+q^{-4}),
\nonumber\\
\chi_{{\bf 4928}'}(q)&=
1904(q+q^{-1})+560(q^3+q^{-3}),
\nonumber\\
\chi_{\bf 8008}(q)&=
3740+1814(q^2+q^{-2})+320(q^4+q^{-4}),
\nonumber\\
\chi_{\bf 8085}(q)&=
3465+2100(q^2+q^{-2})+210(q^4+q^{-4}),
\nonumber\\
\chi_{\bf 8800}(q)&=
3200(q+q^{-1})+1200(q^3+q^{-3}),
\nonumber\\
\chi_{\bf 9152}(q)&=
3520(q+q^{-1})+880(q^3+q^{-3})+160(q^5+q^{-5})+16(q^7+q^{-7}),
\nonumber\\
\chi_{\bf 9504}(q)&=
3312(q+q^{-1})+1440(q^3+q^{-3}),
\nonumber\\
\chi_{\bf 11011}(q)&=
5005+2002(q^2+q^{-2})+715(q^4+q^{-4})+220(q^6+q^{-6})+55(q^8+q^{-8})
\nonumber\\[-4pt]&\quad
+10(q^{10}+q^{-10})+(q^{12}+q^{-12}),
\nonumber\\
\chi_{\bf 11088}(q)&=
5280+2310(q^2+q^{-2})+540(q^4+q^{-4})+54(q^6+q^{-6}),
\nonumber\\
\chi_{{\bf 11088}'}(q)&=
5368+2200(q^2+q^{-2})+550(q^4+q^{-4})+100(q^6+q^{-6})+10(q^8+q^{-8}),
\nonumber\\
\chi_{\bf 11232}(q)&=
5292+2475(q^2+q^{-2})+450(q^4+q^{-4})+45(q^6+q^{-6}),
\nonumber\\
\chi_{\bf 13728}(q)&=
5280(q+q^{-1})+1440(q^3+q^{-3})+144(q^5+q^{-5}).
\end{align}

\section*{Acknowledgements}
We are grateful to Yasuyuki Hatsuda, Sung-Soo Kim, Kimyeong Lee, Takuya Matsumoto, Masatoshi Noumi, Soichi Okada, Kazumi Okuyama, Masato Taki, Akihiro Tsuchiya, Futoshi Yagi, Shintarou Yanagida and especially Yasuhiko Yamada for valuable discussions.
We would like to thank Kobe university, department of mathematics for the kind hospitality where the final part of this work was performed.
We are indebted to the authors of \cite{HHO} for sharing their WKB data up to ${\mathcal O}(k^{29})$ with us.
The work of S.M. is supported by JSPS Grant-in-Aid for Scientific Research (C) \# 26400245.

\end{document}